\documentclass{JFM-FLM_Au}
\usepackage{subcaption}

\usepackage[markup=underlined,authormarkup=superscript]{changes}

\usepackage{xcolor}

\newcommand{\davg}[1]{\left[\kern-0.15em\left[ #1 \right]\kern-0.15em\right]_z}

\lefttitle{G. Li et al.}
\righttitle{Heat transfer enhancement in polymer solutions}

\title{Significant heat transfer enhancement via polymer additives in two-dimensional sheared convection}

\author{ Guanhan Li \aff{1} \corresp{\email{gl523@cam.ac.uk}},\, Lu Zhu\aff{1} \corresp{\email{lz447@cam.ac.uk}}\,\, \& \, Rich R. Kerswell\aff{1}}

\affiliation{\aff{1}Department of Applied Mathematics and Theoretical Physics, University of Cambridge, Wilberforce Road, Cambridge CB3 0WA, UK}

\begin{document}
\maketitle

\begin{abstract}

Heat dissipation is critical in modern engineering and polymer additives offer a potential route to improve fluid-based cooling. We study elasticity-enhanced heat transfer in two-dimensional, thermally stratified Poiseuille flow.
At Reynolds numbers $Re \lesssim 1000$, we observe two linearly unstable modes: the recently identified elasticity-induced centre mode (Khalid \textit{et al.}, \textit{J. Fluid Mech.}\@\ \textbf{915}, 2021) and the classical buoyancy-driven convective mode (Kelly, \textit{Adv. Appl. Mech.}\@\nobreakspace\textbf{31}, 35--112, 1994).
Direct numerical simulations show that the centre mode develops into a nonlinear `arrowhead' state but yields negligible heat transfer enhancement (typically $\approx 0.03\%$ above the conductive state). By contrast, polymers can enhance the heat flux associated with the convective mode by up to  $1100\%$.
The nonlinear convective-mode states are periodic orbits or travelling waves, dominated by hook-like polymer-stress structures that can attach to channel walls.
Both unattached and attached hooks act as `polymer walls', steering flow along rather than across them. Unattached hooks deflect fluid vertically around them in the channel interior and reduce flow rate, whereas wall-attached hooks travel more slowly, reorganising the flow into strong counter-rotating rolls and triggering extreme enhancement.
The elasto-buoyant nature of these states is confirmed by perturbation kinetic energy budgets, showing that polymer and buoyancy sustain the states synergistically.
Wall-attached hooks enable rapid thermal equilibration but impose a large hydraulic penalty, making them suitable for process streams requiring fast temperature adjustment. Unattached hooks provide a more thermally efficient regime for heat transport. 
These results highlight the potential of elastic fluids for future heat transfer enhancement technologies.
\end{abstract}


%
%
\section{Introduction}
\label{sec:intro}

Modern electrical devices often demand efficient cooling systems to dissipate heat and maintain optimal operating conditions \citep{zhang_review_2021}. For example, driven by significant progress in lithography and advanced packaging, the typical heat generated by modern high-performance electronics requires cooling heat fluxes exceeding $10^3\ \mathrm{W/cm^2}$ \citep{wu_cooling_2025}. The present primary cooling system remains air-based, but the next generation focuses on fluid-flow based systems, underscoring the importance of hydrodynamics in contemporary thermal management \citep{konovalov_recent_2023, wu_cooling_2025}.

In many practical cooling configurations, a working fluid passes through a channel where one wall is heated and the opposite wall is much cooler, producing a thermally stratified layer. A canonical framework that captures this interplay of forced and buoyancy-driven motions is the Rayleigh–Bénard–Poiseuille (RBP) flow, in which a pressure-driven stream in a horizontal channel is subjected to a transverse temperature gradient.
Instabilities in this configuration manifest through two primary pathways: shear mechanisms dependent on the Reynolds number $Re$ that provoke Tollmien–Schlichting waves \citep{schmid_stability_2001}, and buoyancy-driven modes governed by the Rayleigh number $Ra$ or the Richardson number $Ri$ that result in the formation of convective rolls \citep{kelly_onset_1994, nicolas_revue_2002}.
However, in many practical applications, such as microchannels in solar units or electronic cooling chips \citep{sharma_energy_2015, khalili_hybrid_2023}, $Re$ typically remains of the order $\mathcal{O}(10^2)$, a regime where only buoyancy instability is triggered. Consequently, leveraging buoyancy instability to optimise thermal performance has become an active area of research. \cite{taher_experimental_2018} experimentally demonstrated that buoyancy-induced secondary flows significantly enhance heat transfer compared to pure forced convection. Expanding on this work, \cite{taher_poiseuille-rayleigh-benard_2021} combined experiments with simulations to provide a detailed map of flow pattern evolution throughout the entire channel: from the entrance to the later turbulent regions. Furthermore, \cite{zhang_three-dimensional_2023} employed 3D numerical simulations to investigate how Rayleigh number $Ra$, Reynolds number $Re$ and lateral confinement affect the stability limits and transitions between longitudinal and transverse roll structures in mixed convection. Beyond passive mechanisms, active control strategies to further increase heat transfer in RBP flows \citep{yan_enhanced_2023, yan_study_2026} have also been investigated.
However, continued miniaturisation reduces channel dimensions and hence the Rayleigh number, which scales with the cube of the channel height. In microscale channels, $Ra$ can therefore fall below the critical threshold for buoyancy-driven instability, leaving the flow laminar and eliminating buoyancy-induced mixing.

One potential strategy to overcome this suppression of instability and hence mixing is the addition of long-chain polymers to the working fluid. The stretching and relaxation of these polymer chains give rise to elastic stresses, which can fundamentally modify flow stability~\citep{steinberg2021elastic,Datta22,dubief2023elasto}. In channels featuring curved geometries, such as wavy or serpentine paths or those containing internal obstacles, the well-known hoop stress polymer instability can lead to elastic turbulence \citep{groisman_elastic_2000}. 
To date, the literature on elasticity-induced heat-transfer enhancement (HTE) has largely focused on this phenomenon. For instance, \cite{abed_experimental_2016} experimentally observed that heat transfer increased by up to 200\% for low polymer concentrations and up to 380\% for higher concentrations compared to equivalent Newtonian flows in a serpentine channel. Similarly, \cite{roy_viscoelastic_2024} numerically demonstrated that heat transfer can reach approximately twice that of its Newtonian counterpart in channels with obstacles. Other experimental \citep{traore_efficient_2015, whalley_enhancing_2015, yang_experimental_2020} and numerical \citep{li_numerical_2017, garg_enhanced_2024} studies have confirmed this HTE phenomenon induced by elastic turbulence. 

Conversely, in straight-channel flows, some studies have reported a reduction in heat transfer in Rayleigh-Bénard convection \citep{cai_polymer_2019, xu_direct_2025}, whereas others have observed both enhancement and suppression, with the reported enhancement remaining modest (e.g. about 2\% in \citealt{wang_pattern_2023} and about 10\% in \citealt{dubief_heat_2020}). Recently, it has been found that other viscoelastic instabilities, such as the centre mode \citep{khalid_centre-mode_2021} and polymer diffusive instability (PDI) \citep{beneitez_polymer_2023}, can be triggered in a straight channel. However, the effectiveness of the resulting motions for heat flux has yet to be explored. This study represents a first attempt to address this issue, specifically focusing on cases where the centre mode is excited.

In this paper, we will conduct linear stability analysis and direct numerical simulations to investigate HTE in viscoelastic, unstably stratified (RBP flow) and stably stratified shear flow. To the best of our knowledge, previous studies have only focused on the stability analysis of similar systems, and the HTE effect has not been previously investigated in the literature as this requires a finite-amplitude manifestation of the instability to assess. An early investigation of viscoelastic RBP flow \citep{hirata_convective_2015} focused on the convective and absolute instabilities by linear stability analysis, finding that while fluid elasticity acts to stabilise transverse rolls in the presence of throughflow, it significantly accelerates the transition to absolute instability. \cite{yao_effects_2024} also conducted linear stability analysis and identified three distinct instability regimes governed by shear intensity (quantified by $Re$), revealing a non-monotonic dependence where viscoelasticity acts to destabilise the flow at low and high $Re$ but plays a stabilising role at moderate $Re$. In contrast to these modal analyses, \cite{yao_non-modal_2025} focused on the linear non-modal dynamics of similar systems and uncovered a substantial transient energy growth driven by the coupling between polymeric stresses and temperature gradients. Furthermore, \cite{kamboj_exacerbation_2025} considered the effect of viscous heating, highlighting that the buoyancy induced by internal viscous dissipation can dramatically exacerbate purely elastic instabilities by lowering the critical Weissenberg number $W$.

The rest of this paper is organised as follows. 
\S\ref{sec:problem_formulation} 
formulates the physical problem and its governing equations. 
\S\ref{sec:lsa} is dedicated to linear stability analysis and a discussion of the 
fastest-growing modes.
In \S\ref{sec:dns}, we conduct the direct numerical simulations (DNS) to investigate the evolution of those unstable modes, including the numerical setup (\S\ref{sec:dns_setup}), a characterisation of representative flow cases (\S\ref{sec:six_cases}), and a parametric study of the Reynolds and Weissenberg numbers (\S\ref{sec:parametric_re_w}).
We then investigate the connection between the flow field and any HTE effects in \S\ref{sec:flow_field}, and explore the self-sustaining mechanism of the nonlinear resulting states through a perturbation kinetic energy budget analysis in \S\ref{sec:energy_budget}.
In \S\ref{sec:engineering}, we evaluate the thermal-hydraulic efficiency from an engineering perspective.
Finally, concluding 
remarks are provided in \S\ref{sec:conclusions}.

%
%
\section{Problem formulation}
\label{sec:problem_formulation}

%
%
\subsection{Governing equations}

%
%
\begin{figure}
  \centerline{\includegraphics[width=0.6\linewidth]{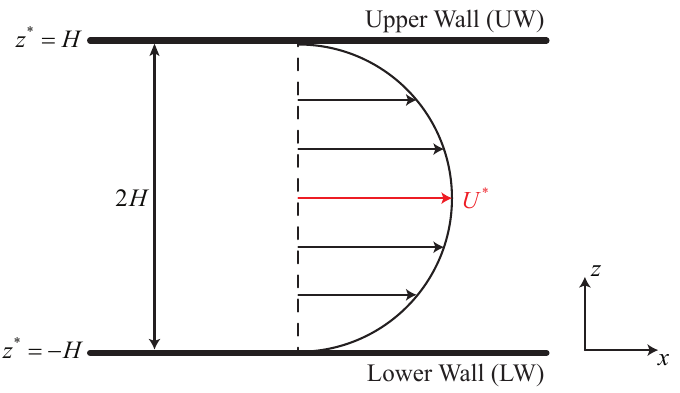}}
  \caption{Schematic of the Poiseuille channel flow.}
  \label{fig:pdf}
\end{figure}

We consider pressure-driven flow of an incompressible viscoelastic fluid in a channel with two walls at $z = \pm H$, as shown in Figure~\ref{fig:pdf}. The dimensionless continuity, Navier-Stokes, Oldroyd-B and density transport equations under the Boussinesq approximation can be written as:
\begin{equation}
\nabla\cdot\boldsymbol{u} = 0,
\label{eq:vec_continuity}
\end{equation}
\begin{equation}
\frac{\partial \boldsymbol{u}}{\partial t}
+ \boldsymbol{u}\cdot\nabla\boldsymbol{u}
= -\nabla p
+ \frac{1-\beta}{Re}\nabla\cdot\boldsymbol{\tau}
+ \frac{\beta}{Re}\nabla^2\boldsymbol{u}
+ Ri\rho\boldsymbol{\hat z} \,
+ \frac{2}{Re}\boldsymbol{\hat x},
\label{eq:vec_ns}
\end{equation}
\begin{equation}
\frac{{\partial {\boldsymbol{\tau }}}}{{\partial t}} + {\boldsymbol{u}} \cdot \nabla {\boldsymbol{\tau }} - {(\nabla {\boldsymbol{u}})^\mathrm{T}} \cdot {\boldsymbol{\tau }} - {\boldsymbol{\tau }} \cdot \nabla {\boldsymbol{u}} + \frac{1}{{{W}{\mkern 1mu} }}{\boldsymbol{\tau }} = \frac{1}{{{W}{\mkern 1mu} }}\left[ {\nabla {\boldsymbol{u}} + {{(\nabla {\boldsymbol{u}})}^\mathrm{T}}} \right] + {\mkern 1mu} \frac{1}{{{\mathop{Re}\nolimits} Sc}} {\mkern 1mu} {\nabla ^2}{\boldsymbol{\tau }},
\label{eq:vec_oldroyd}
\end{equation}
\begin{equation}
{\mkern 1mu} \frac{{\partial \rho }}{{\partial t}} + {\boldsymbol{u}} \cdot \nabla \rho  = \frac{1}{{{Pr}{\mkern 1mu} {Re}}}{\nabla ^2}\rho.
\label{eq:vec_thermal}
\end{equation}
These equations have been non-dimensionalised by the centreline speed of the base Poiseuille flow solution $U^*$ (see (\ref{Ubase}) below), the half-height of the channel $H$, and the densities imposed at the isothermal walls ($\rho_{UW}^*$ is the enforced density at the upper wall and $\rho_{LW}^*$ at the lower wall and  $\rho_{UW}^*-\rho_{LW}^* \ll \rho_{LW}^*$ for the Boussinesq approximation to hold). The 2D velocity field $\boldsymbol{u}$ is therefore measured in units of $U^*$, the polymer stress tensor $\boldsymbol{\tau}$ in units of $U^*/H$ and $\rho$, the  density perturbation away from $\rho_{LW}^*$ by $\rho_{UW}^*-\rho_{LW}^*$. 
The polymer stress diffusion coefficient $\delta$, the thermal diffusivity $\kappa$, the relaxation time $\lambda$, the total dynamic viscosity $\mu = {\mu_s} + {\mu_p}$ (where ${\mu_s}$ and ${\mu_p}$ are the solvent and polymer dynamic viscosities, respectively) and the total kinematic viscosity $\nu=\mu/\rho_{LW}^*$ are dimensional material properties that enter the governing equations through the following dimensionless parameters:
$$
\beta := \frac{\mu_s}{\mu}, 
\qquad Sc := \frac{\nu }{\delta },
\qquad Re := \frac{\rho_{LW}^* U^* H}{\mu} = \frac{U^* H}{\nu},
$$
$$
Ri :=  \frac{{g{\mkern 1mu} {\mkern 1mu} H\left( {\rho_{LW}^*-\rho_{UW}^*} \right)}}{{{\rho _{LW}^*}{U^*}^2}}, 
\qquad Pr := \frac{\nu}{\kappa}, 
\qquad W := \frac{U^* \lambda}{H}.
$$
Here  $\beta$ is the solvent viscosity ratio quantifying polymer concentration in the fluid ($0 \le \beta \le 1$), $Sc$ is the Schmidt number, $Re$ is the Reynolds number and is taken to be on the order of $\mathcal{O}(10^2)$ to match applications in the microchannels \citep{sharma_energy_2015, khalili_hybrid_2023}, $Ri$ is the Richardson number ($Ri<0$ indicates unstable stratification), $Pr$ is the Prandtl number, which is fixed to $Pr = 7$ to represent water at room temperature, and $W$ is the Weissenberg number. 
Under this non-dimensionalisation, the imposed pressure gradient per unit mass is $2/Re\, {\boldsymbol{ \hat{x} } }$ and is therefore fixed in time for a given $Re$ whereas the mass flux is variable and dependent  on the exact  flow state.

For further convenience, we also define Rayleigh number $Ra$ to match previous studies \citep[e.g.][]{gage_stability_1968}:
\begin{equation}
    Ra := \frac{{\alpha g (2H)^3 \left( {\theta _{LW}^* - \theta _{UW}^*} \right)}}{{\kappa \nu }}=- 8Ri{{\mathop{Re}\nolimits} ^2}Pr,
    \label{eq:Rayleigh_number}
\end{equation}
where $\alpha (\theta_{LW}^*-\theta_{UW}^*)=\rho_{UW}^*-\rho_{LW}^*$ with $\alpha$ the coefficient of thermal expansion and $\theta _{LW}^*$ and $\theta _{UW}^*$ being the  temperatures at the lower and upper walls respectively.

The Oldroyd-B model is used here to model the polymer dynamics given its relative mathematical simplicity (e.g. the base state can be written down explicitly - see (\ref{Ubase}) and (\ref{eq:basestate}) below) and the small number of extra parameters it introduces ($\beta$ and $W$ with $Sc$ added to stabilise numerical solutions). The model is also derivable from a microscopic description, where each polymer molecule is modelled as two  masses connected by a Hookean spring (a dumbbell), but  it has known deficiencies - e.g. the dumbbells can be infinitely extended (for recent discussions see \cite{Shaqfeh21,Castillo22}). Repeating the computations reported below with a more sophisticated model (see Table~1 in \cite{Datta22} for more sophisticated continuum models and e.g. \cite{Serafini22} for hybrid models) would clearly be very valuable.

%
%
\subsection{Boundary conditions}

In Poiseuille flow, the streamwise direction $x\in[0,L_x]$ is taken as periodic and the wall-normal direction $z\in[-1,1]$ is bounded by no-slip, impermeable walls at $z=\pm1$, where $\boldsymbol{u}=(0,0)^\mathrm{T}$. No boundary conditions are required for the pressure whereas Dirichlet conditions are imposed on the density field: $\rho(-1)=0$ and $\rho(1)=1$ to indicate fixed temperature boundaries. We set $Sc=\infty$ in the linear stability analysis for simplicity (so no polymer stress boundary conditions are needed) and a sufficiently large but finite $Sc$ in the DNS for numerical stability. Therefore, extra boundary conditions on the polymer stress tensor $\boldsymbol{\tau}$ are required as detailed in the following sections.

%
%
\subsection{Base state}
\label{sec:base_state}

To perform the linear stability analysis, we decompose each field into the sum of its base‐state value (uppercase letter) and a perturbation (lowercase letter with superscript $'$):
\begin{equation}
\begin{array}{l}
\boldsymbol{u}\left( {x,z,t} \right) = {\boldsymbol{U}}\left( {z} \right) + \boldsymbol{u}'\left( {x,z,t} \right),\\
p \left( {x,z,t} \right) = {P}\left( {z} \right) + p'\left( {x,z,t} \right),\\
\boldsymbol{\tau} \left( {x,z,t} \right) = {\boldsymbol{T}}\left( {z} \right) + \boldsymbol{\tau} '\left( {x,z,t} \right),\\
\rho \left( {x,z,t} \right) = {R}\left( {z} \right) + \rho'\left( {x,z,t} \right).
\end{array}
\label{eq:base_state_decomposition}
\end{equation}
The base state  taken throughout is fully developed Poiseuille flow with a linearly stratified density field and a concomitant polymer stress field at $Sc=\infty$, i.e.,
\begin{align}
{{\boldsymbol{U}}}\left( z \right) 
& = 
\left[ {\begin{array}{*{20}{c}}
{{U}}\\0
\end{array}} \right] 
= \left[ \begin{array}{c}
1 - {z^2}\\0
\end{array} \right],  \label{Ubase}\\
    R = \tfrac{1}{2}(1+z),
    \qquad 
    \boldsymbol{T} &= \left[ {\begin{array}{cc}
    {{T_{xx}}}&{{T_{xz}}}\\
    {{T_{xz}}}&{{T_{zz}}}
    \end{array}} \right] = \left[ {\begin{array}{cc}
    {8W{z^2}}&{ - 2z}\\
    { - 2z}&0
    \end{array}} \right].   \label{eq:basestate}
\end{align}

%
%
\section{Linear stability analysis}
\label{sec:lsa}

In this section, we carry out a linear stability analysis to obtain a preliminary overview of the unstable modes arising within the chosen region of parameter space. In the $W$--$Ri$ parameter space, two unstable modes are identified: a convective mode and an elastic centre mode. We then examine the effect of $W$ on the convective mode and that of $Ri$ on the elastic centre mode, in order to clarify how elasticity and stratification respectively modify the structure and destabilisation characteristics of these modes.

%
%
\subsection{Methodology}
\label{sec:method_lsa}

Due to the steadiness of the base state and its translational invariance in $x$, the perturbation can be Fourier-decomposed as
\begin{equation}
( \boldsymbol{u}',p',\rho',\boldsymbol{\tau}' )(x,z,t) 
= 
\left(
\left[ \begin{array}{l}
\tilde u\left( z \right)\\
\tilde w\left( z \right)
\end{array} \right],
\tilde p(z), 
\tilde \rho (z), 
\left[ {\begin{array}{*{20}{c}}
{{{\tilde \tau }_{xx}}\left( z \right)}&{{{\tilde \tau }_{xz}}\left( z \right)}\\
{{{\tilde \tau }_{xz}}\left( z \right)}&{{{\tilde \tau }_{zz}}\left( z \right)}
\end{array}} \right] 
\right)e^{ik\left( {x - ct} \right)}
\label{eq:normal_mode}
\end{equation}
with the objective of studying how the eigenvalue $c\in\mathbb{C}$ varies with $k \in \mathbb{R}$. The system is unstable if $c_i>0$ for some $k$  (where the subscript $i$ denotes the imaginary part) with $\omega_i = kc_i$ being the growth rate of the linear instability. This normal mode ansatz leads to  the following linear equations:
\begin{equation} 
ikc  \left[ \begin{array}{c}
Re \,\tilde u \\
  \\[7pt]
Re \, \tilde w \\
\\[9pt]
W \,\tilde \tau_{xx}\\
\\[12pt]
W \,\tilde \tau_{xz}\\
\\[12pt]
W \,\tilde \tau_{zz}\\
\\[15pt]
Pr Re \,\tilde \rho\\
\\[7pt]
0
\end{array}
\right]
=
\left[ \begin{array}{l}
\left( {Re}ik{U} + \beta k^{2} - \beta D^{2} \right)\tilde u + \left( {Re}D{U} \right)\tilde w + ik{Re}\tilde p\\
+\left[ - (1-\beta)ik \right]\tilde \tau_{xx} + \left[ - (1-\beta)D \right]\tilde \tau_{xz}\\[9pt]
\left( {Re}ik{U} - \beta D^{2} + \beta k^{2} \right)\tilde w + {Re}D\tilde p + \left[ - (1-\beta)ik \right]\tilde \tau_{xz}\\
+ \left[ - (1-\beta)D \right]\tilde \tau_{zz} - {Re}\,{Ri}\,\tilde \rho\\[9pt]
\bigl(-2Wik\,T_{xx} - 2WT_{xz}D - 2ik\bigr)\tilde u + WDT_{xx}\,\tilde w \\
+ \Bigl(iWkU + 1 + W/(ReSc) k^{2} - W/(ReSc) D^{2}\Bigr)\tilde \tau_{xx} - 2WD\,U\,\tilde \tau_{xz}\\[7pt]
\bigl(-WT_{zz}D - D\bigr)\tilde u + \bigl(WDT_{xz} - iWkT_{xx} - ik\bigr)\tilde w+\\
\Bigl(1 + W/(ReSc) k^{2} - W/(ReSc) D^{2} + iWkU\Bigr)\tilde \tau_{xz} - WD\,U\,\tilde \tau_{zz}\\[7pt]
\bigl(WDT_{zz} - 2iWkT_{xz} - 2WT_{zz}D - 2D\bigr)\tilde w+\\
\Bigl(1 + W/(ReSc) k^{2} - W/(ReSc) D^{2} + iWkU\Bigr)\tilde \tau_{zz}\\[7pt]
Pr\,Re\,DR\,\tilde w + \Bigl(iPr\,Re\,kU - (D^{2} - k^{2})\Bigr)\tilde\rho \\[7pt]
\\
ik\,\tilde u + D\,\tilde w  
\end{array}
\right],
\label{eq:final_combined_lsa}
\end{equation}
where $D := d/dz$. Again we remark that the polymer stress diffusion term $\nabla^2\boldsymbol{\tau}$ is removed in the linear stability analysis so no boundary conditions are required for $\boldsymbol{\tau}'$. The velocity perturbation satisfies no-slip conditions, $\boldsymbol{u}'(\pm1)=(0,0)^\mathrm{T}$, the density perturbation satisfies $\rho'(\pm1)=0$, and no boundary conditions are needed on the pressure perturbation $p'$.

We discretise the linearised system (\ref{eq:final_combined_lsa}) by a Chebyshev–Galerkin spectral method on the interval $z\in[-1,1]$. The unknown variables with no boundary conditions, namely $\tilde{p}$, $\tilde{\tau}_{xx}$, $\tilde{\tau}_{xz}$, and $\tilde{\tau}_{zz}$, are expanded using the standard Chebyshev polynomials $T_n(z):=\cos( n \cos^{-1}(z))$.
The remaining fields $\tilde{u}, \tilde{w},$ and $\tilde{\rho}$ are expanded using the basis functions $\phi_n(z):= T_{n+2}(z) - T_{n}(z)$ which automatically satisfy homogeneous Dirichlet boundary conditions. This linear system then becomes a generalised eigenvalue problem of the following form:
\begin{equation}
\boldsymbol{A}\,\boldsymbol{x} = c\,\boldsymbol{B}\,\boldsymbol{x},
\end{equation}
where $\boldsymbol{A}$ and $\boldsymbol{B}$ are ${\mathbb C^{7N \times 7N}}$ matrices  and ${\boldsymbol{x}} = {\left[ {\tilde u,\tilde w,\tilde p,{{\tilde \tau }_{xx}},{{\tilde \tau }_{xz}},{{\tilde \tau }_{zz}},\tilde \rho } \right]^\mathrm{T}}$ contains the coefficients of the spectral expansions at the corresponding collocation points ($N$ is the size of the expansions for each dependent variable).

%
%
\subsection{Results}
\label{sec:lsa_results}

%
%
\begin{figure}
\centering
\begin{subfigure}{0.495\textwidth}
    \centering
    \includegraphics[width=\textwidth]{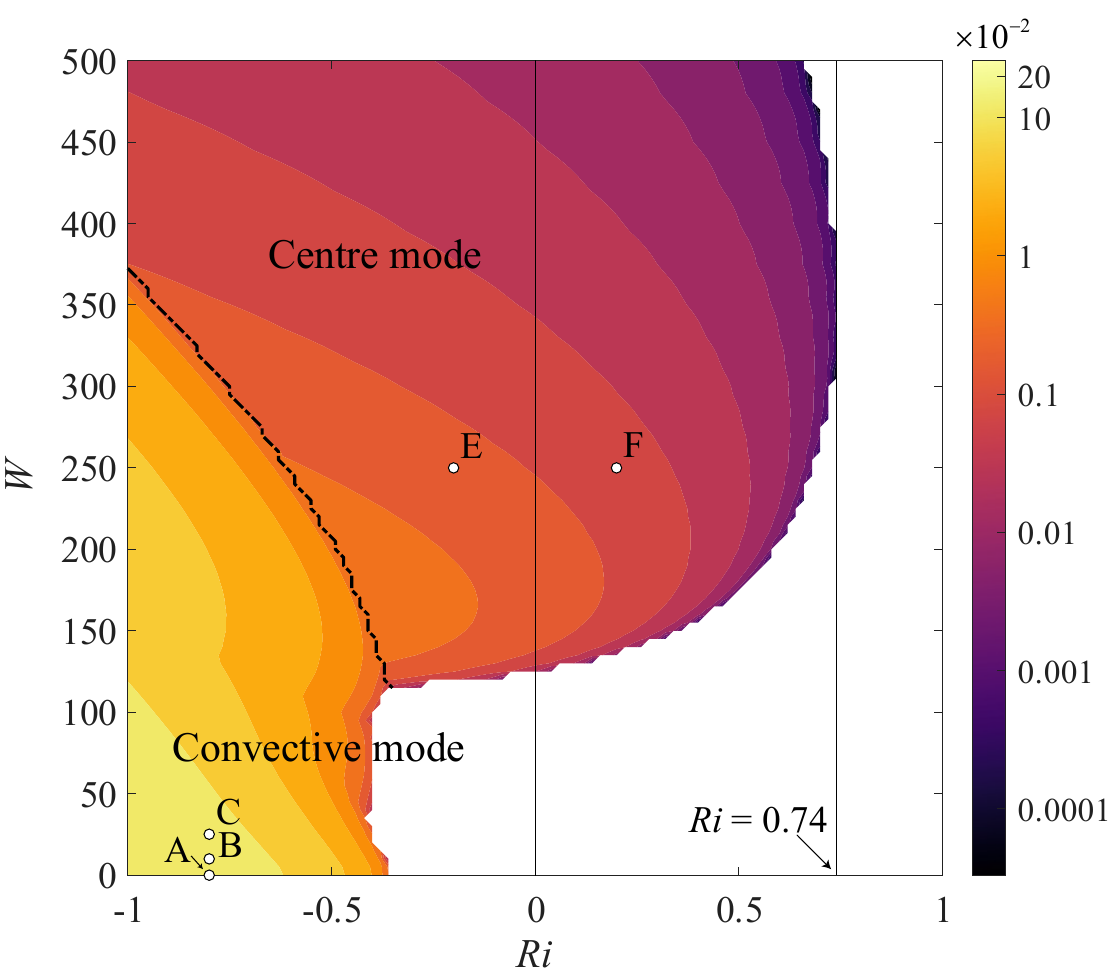}
    \caption{}
\end{subfigure}
\hfill
\begin{subfigure}{0.495\textwidth}
    \centering
     \includegraphics[width=\textwidth]{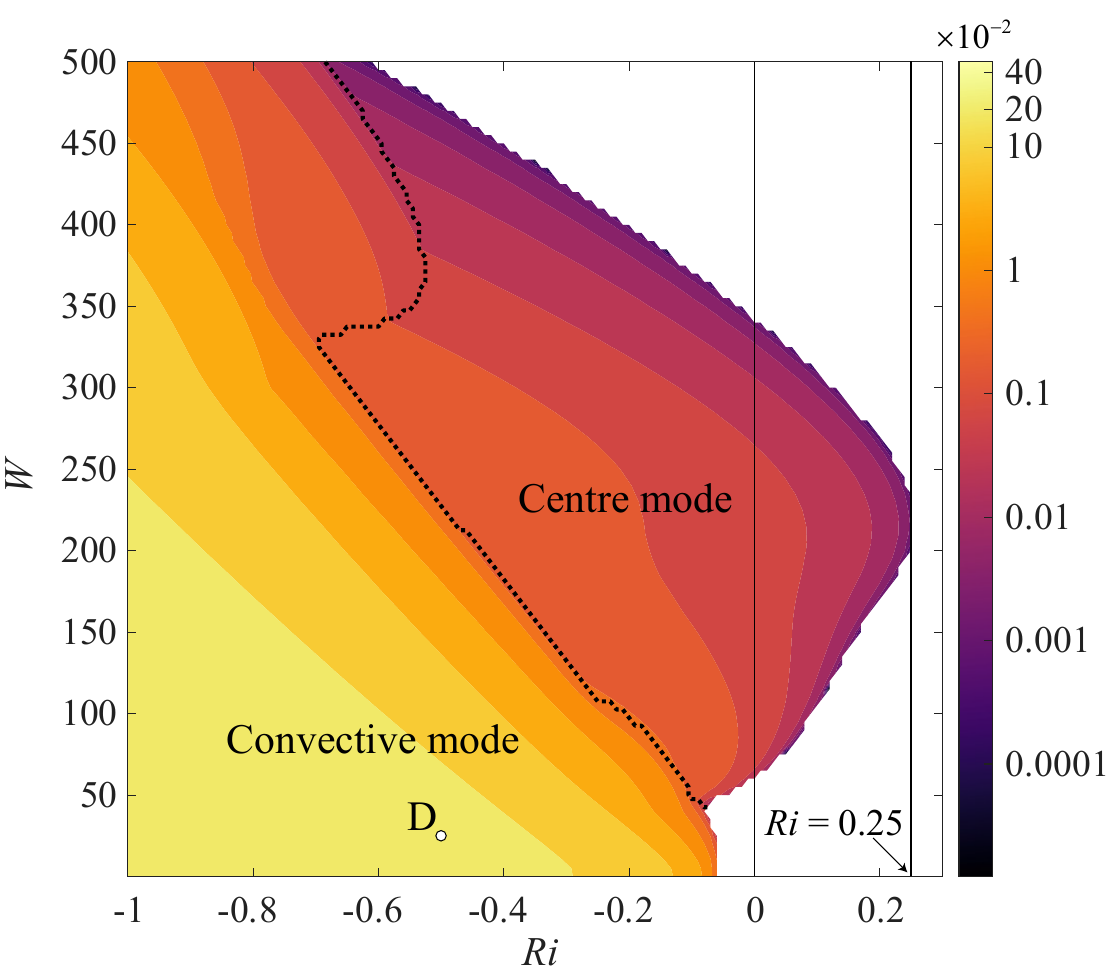}
    \caption{}
\end{subfigure}
\caption{Growth rate of the fastest growing mode at (a) $Re=50$, $\beta=0.98$ and (b) $Re=800$, $\beta=0.8$. The black dashed line marks the transition boundary where the dominant mode switches. In each panel, the left black solid line marks $Ri = 0$, separating the parameter space into unstable (left) and stable (right) stratification. The right black solid line indicates the existence boundary of the centre mode within the stably stratified regime, corresponding to $Ri = 0.74$ (equivalent to $Ra = -1.036 \times 10^5$) in (a) and $Ri=0.25$ ($Ra = -8.96 \times 10^6$) in (b). Markers A, B, C, D, E, F denote representative cases: 
  convective mode A: $(Ri,W)=(-0.8, 0)$; 
  convective mode B: $(Ri,W)=(-0.8, 10)$; 
  convective mode C: $(Ri,W)=(-0.8, 25)$;
  convective mode D: $(Ri,W)=(-0.5, 25)$;
  centre mode E: $(Ri,W)=(-0.2, 250)$;
  centre mode F: $(Ri,W)=(0.2, 250)$.}
\label{fig:lsa_50_800}
\end{figure}

Figure~\ref{fig:lsa_50_800} presents the results of the linear stability analysis in the $Ri$--$W$ plane which highlights the presence of two different modes of instability. The convective mode, corresponding to the transverse rolls indicated by \cite{kelly_onset_1994}, appears only in the unstably stratified region, emerging once $Ri$ falls below a certain threshold. The elastic centre mode found by \cite{khalid_centre-mode_2021} is present in both stable and unstable stratification at large $W$, and triggers instability even when the corresponding Newtonian state is linearly stable. In the following sections, we characterise these two unstable modes in more detail.

%
%
\begin{figure}
\centering

\begin{subfigure}{0.48\textwidth}
    \centering
    \includegraphics[width=\textwidth]{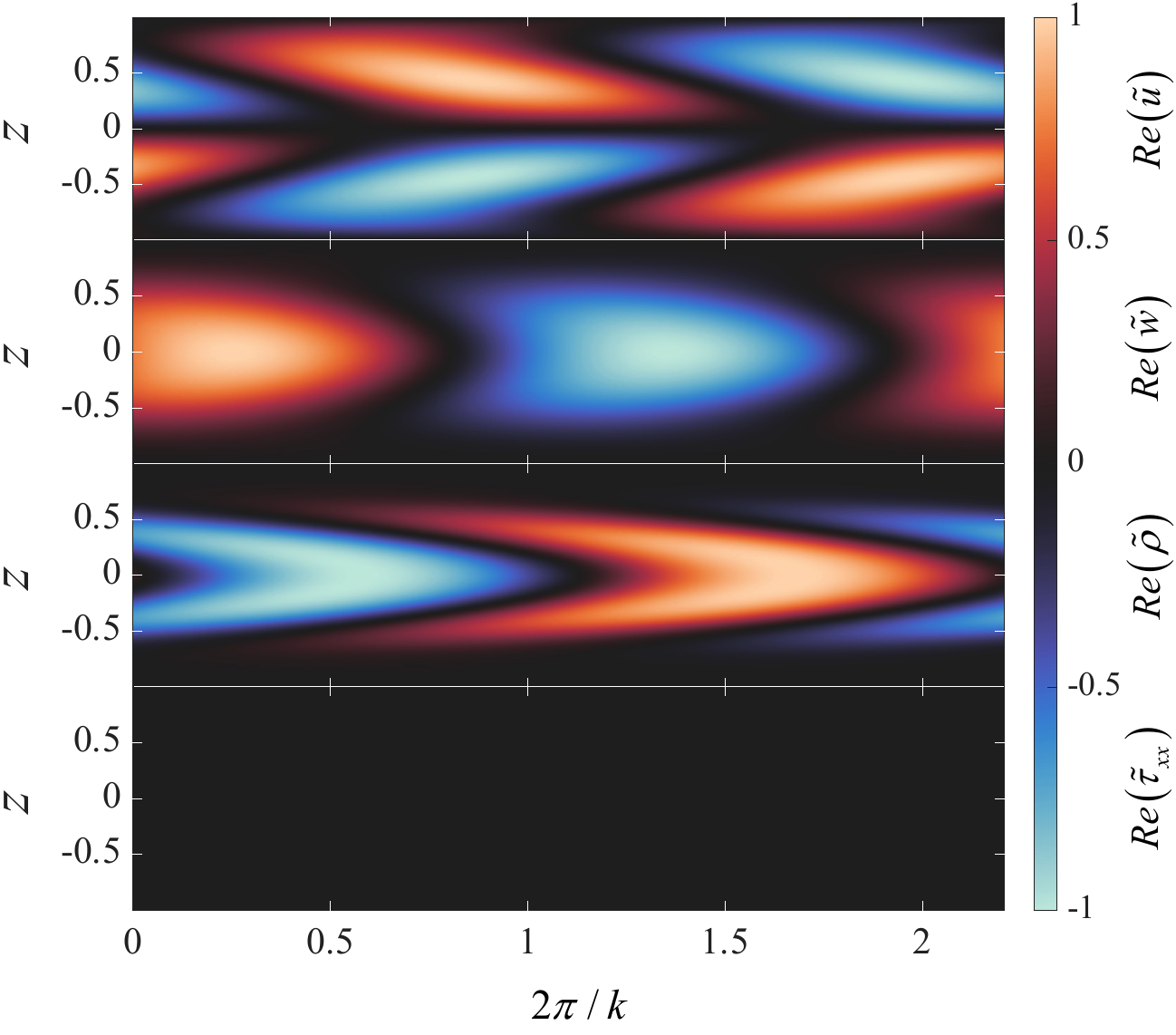}
    \caption{Convective mode A}
\end{subfigure}
\quad
\begin{subfigure}{0.48\textwidth}
    \centering
    \includegraphics[width=\textwidth]{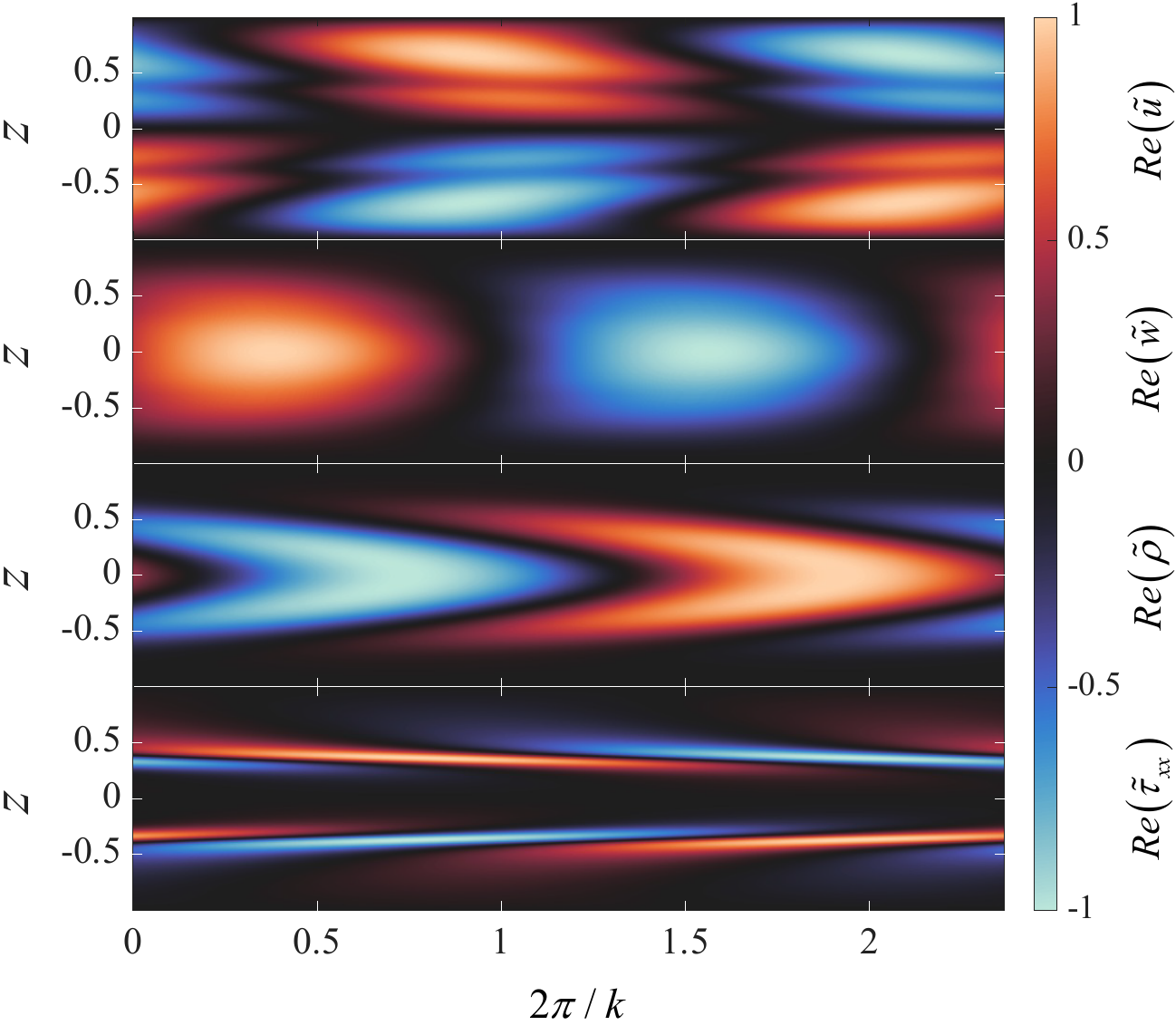}
    \caption{Convective mode C}
\end{subfigure}
\vspace{0.1cm}

\begin{subfigure}{0.48\textwidth}
    \centering
    \includegraphics[width=\textwidth]{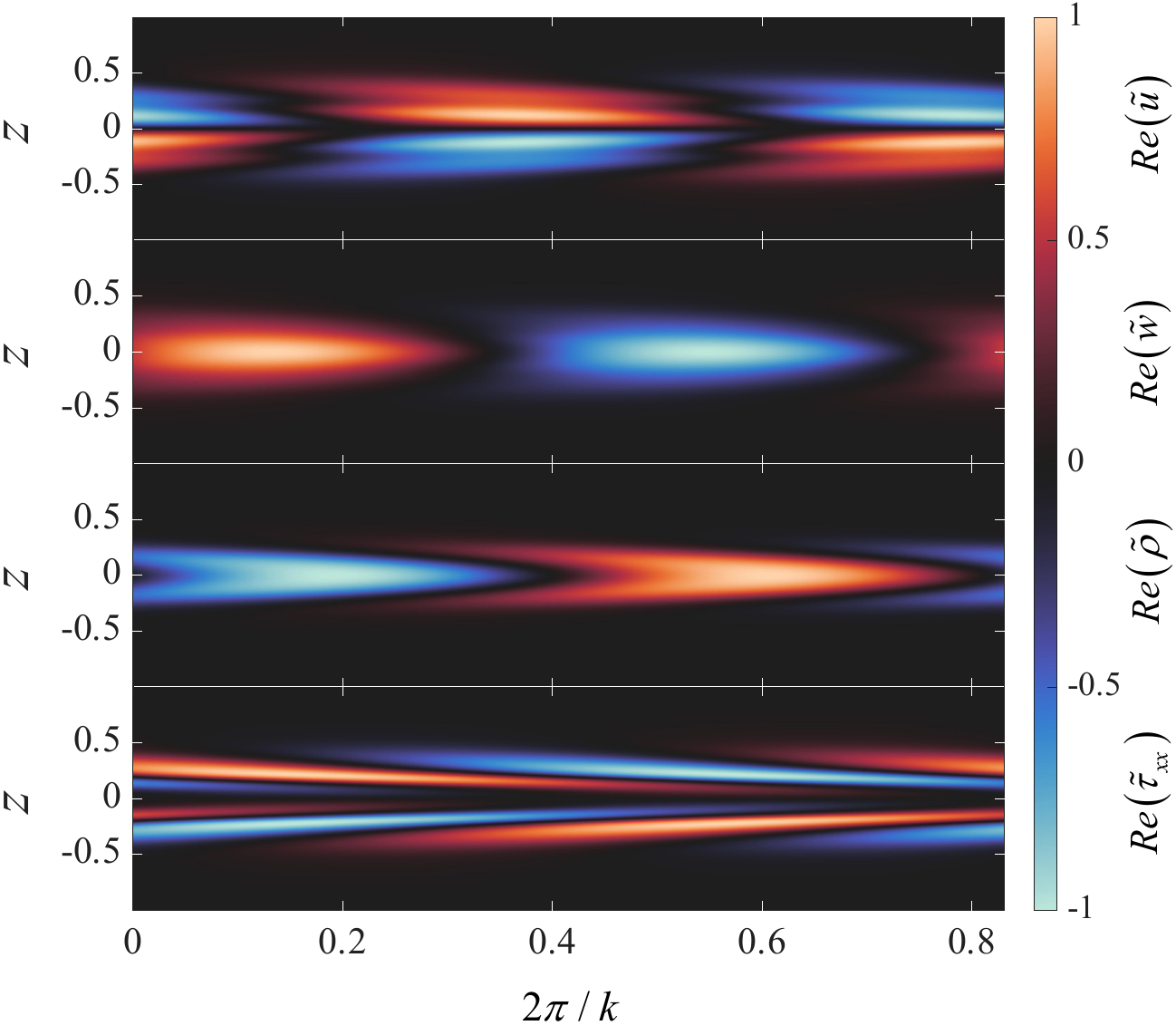}
    \caption{Convective mode D}
\end{subfigure}
\quad
\begin{subfigure}{0.48\textwidth}
    \centering
    \includegraphics[width=\textwidth]{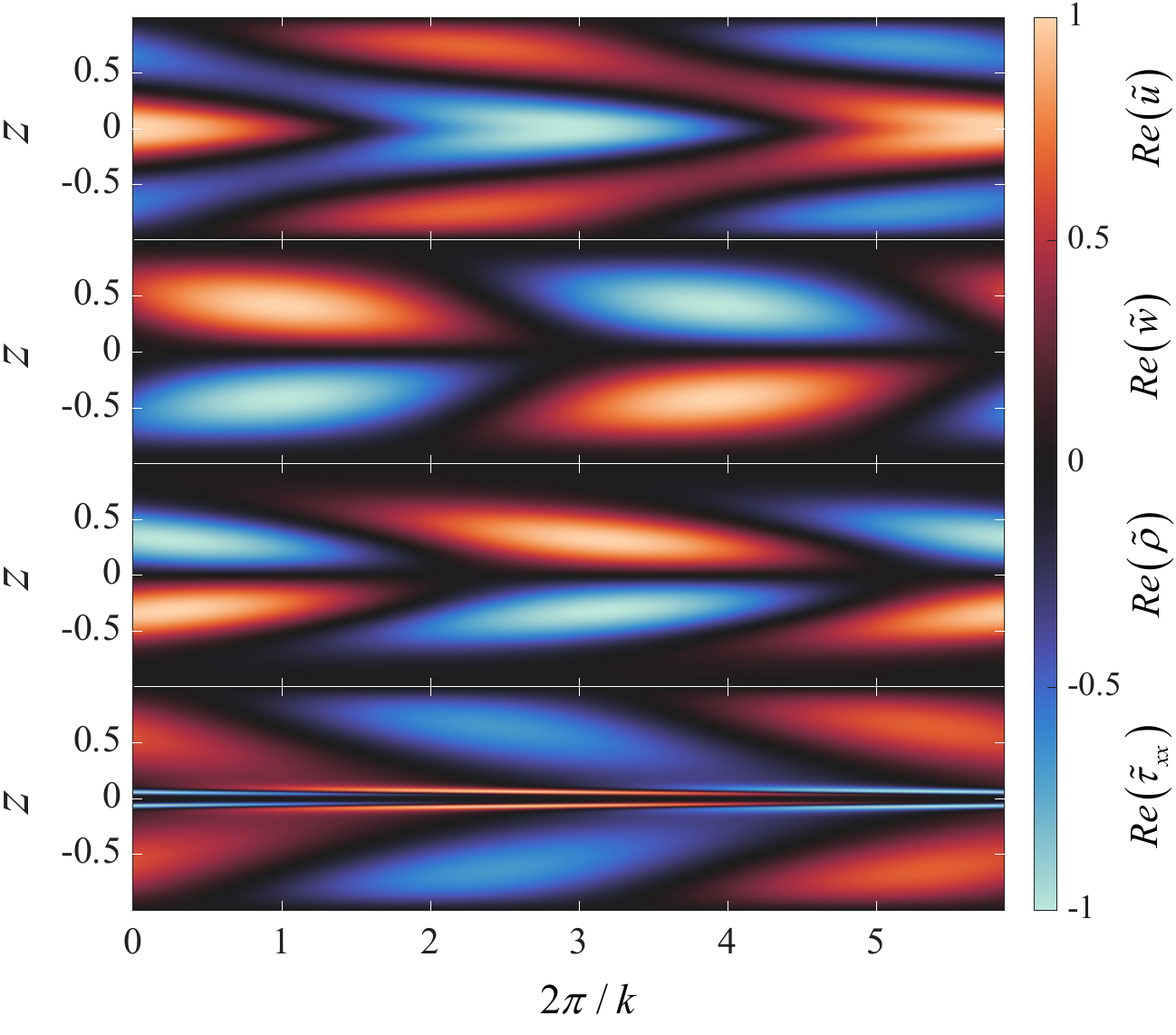}
    \caption{Centre mode F}
\end{subfigure}
\caption{Contours of the eigenfunctions $\tilde u$, $\tilde w$, $\tilde\rho$, and $\tilde\tau_{xx}$ associated with the fastest‐growing modes for the representative cases shown in Figure~\ref{fig:lsa_50_800}. 
For each case, the streamwise velocity $\tilde u$, wall‐normal velocity $\tilde w$, density perturbation $\tilde \rho$, and polymer stress component $\tilde \tau_{xx}$ are displayed in sequence.
The panels correspond to: (a) convective mode~A; (b) convective mode~C; (c) convective mode~D; and (d) centre mode~F.}
\label{fig:eigenfuncs}
\end{figure}

%
%
\begin{figure}
    \captionsetup[subfigure]{labelformat=empty,labelsep=none}
    \centering
    \begin{subfigure}[b]{0.47\textwidth}
        \centering
        \includegraphics[width=\textwidth]{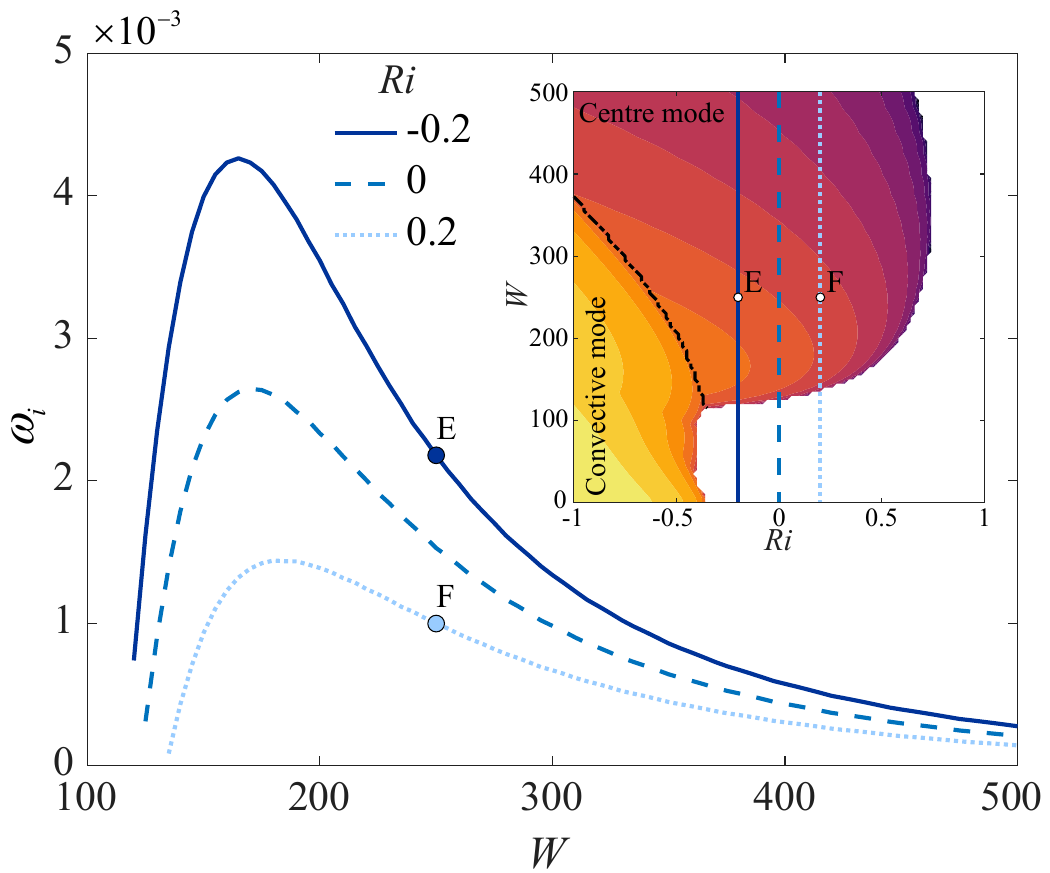}
        \caption{(a)}
    \end{subfigure}
    \quad
    \begin{subfigure}[b]{0.48\textwidth}
        \centering
        \includegraphics[width=\textwidth]{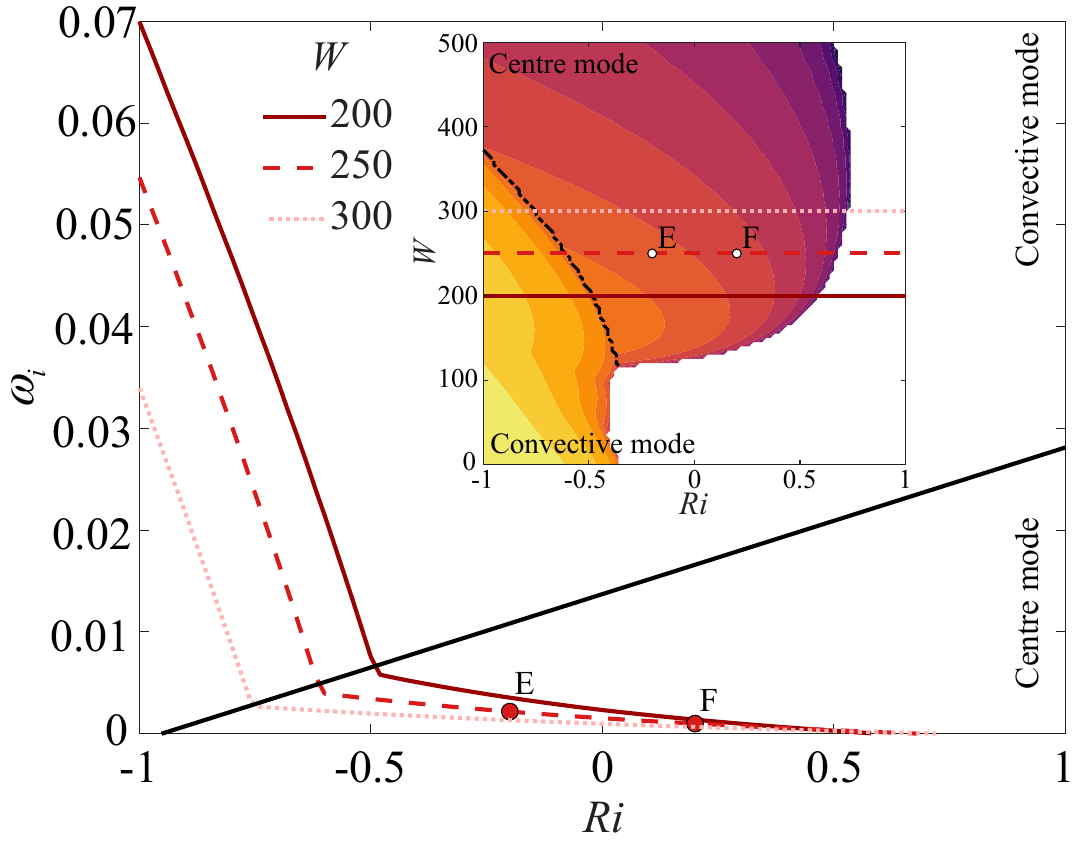}
        \caption{(b)}
    \end{subfigure}
\caption{Variations of the growth rate $\omega_i$ with (a) $W$ and (b) $Ri$ at $Re=50$ and $\beta=0.98$, corresponding to the cases in Figure~\ref{fig:lsa_50_800}(a). The markers E and F denote the centre modes~E and~F labelled in Figure~\ref{fig:lsa_50_800}(a). Note that panel (a) displays only the centre mode, whereas panel (b) includes both convective and centre modes, separated by the black solid line. The insets in both panels show the locations in Figure~\ref{fig:lsa_50_800}(a).}
    \label{fig:gr_vs_W_vs_Ri}
\end{figure}

%
%
\subsubsection{Convective-mode instability}

As shown by the boundary between dominant modes (black dashed line) in Figure~\ref{fig:lsa_50_800}, the convective mode dominates when $Ri$ is sufficiently negative and $W$ is low. The growth rate generally increases as $W$ decreases at fixed $Ri$, with localised non-monotonicity near the stability boundary, and increases monotonically as $Ri$ decreases at fixed $W$.
The dependence of the convective mode on $W$ suggests that the polymers modify the convective flow structure and, consequently, the heat-transfer performance of the convection flow. This will be investigated further in \S\ref{sec:dns}. The eigenfunctions of representative convective modes~A, C, and D are displayed in Figure~\ref{fig:eigenfuncs}(a)--(c) (mode~B is not shown owing to its high similarity to mode~A).
We find for all that $\tilde{u}$ and $\tilde{\tau}_{xx}$ are antisymmetric about the channel centreline, whereas $\tilde{w}$ and $\tilde{\rho}$ are symmetric, as also demonstrated by \cite{yao_effects_2024}. Comparing Figures~\ref{fig:eigenfuncs}(a) and \ref{fig:eigenfuncs}(b), adding polymer additives to the flow cause sharp gradients emerge near $z=\pm 0.5$ in the $\tilde{u}$ eigenfunction. The $\tilde{w}$ eigenfunction also exhibits a gradient, although it is less pronounced; this gradient becomes clearer at higher elasticity. In contrast, the $\tilde{\rho}$ eigenfunction remains nearly unaffected. For the eigenfunctions of convective mode~D (Figure~\ref{fig:eigenfuncs}(c)), the structures are localised near the channel centre with much weaker amplitudes in the surrounding bulk region. This observation is likely attributed to the strong shear effects introduced by the increase in $Re$ \citep{yao_effects_2024}.

%
%
\subsubsection{Centre-mode instability}

The eigenfunction of centre mode~F is shown in Figure~\ref{fig:eigenfuncs}(d) (mode~E is not shown as it is very similar to mode F). The centre mode has $\tilde u$ and $\tilde \tau_{xx}$ symmetric about the channel centreline and $\tilde w$ and $\tilde \rho$ are antisymmetric \citep{khalid_centre-mode_2021}, so the opposite symmetries to the convective mode. Figure~\ref{fig:lsa_50_800} shows that increasing stable stratification suppresses the growth of the centre mode with the centre mode instability  persisting in the stably stratified regime up to $Ri = 0.74$ ($Ra=-1.036\times10^5$ and $Re=50$) in panel (a) and $Ri = 0.25$ ($Ra = -8.96\times10^6$ and $Re=800$) in panel (b) (this threshold goes below 0.25 for higher $Re$ so there is no connection with the Miles-Howard result: \cite{Miles61,Howard61}). 

To investigate the modification in detail, Figure~\ref{fig:gr_vs_W_vs_Ri} presents the variation of the growth rate $\omega_i$ as a function of $W$ for fixed $Ri$ (panel a) and as a function of $Ri$ for fixed $W$ (panel b).
In Figure~\ref{fig:gr_vs_W_vs_Ri}(a), at fixed $Ri$, the growth rate first increases with $W$ when the centre mode emerges, reaches a peak, and subsequently decreases. In Figure~\ref{fig:gr_vs_W_vs_Ri}(b), for fixed $W$, the growth rates of both the convective and centre modes decrease approximately linearly but with distinct slopes: the convective mode exhibits a steep decline, whereas the centre mode decreases more gradually. This indicates, perhaps not surprisingly, that while increasing $Ri$ suppresses instability in both cases, the buoyancy-induced convective mode is more sensitive to stratification.


\section{Direct numerical simulations}
\label{sec:dns}

To investigate the heat transfer associated with the nonlinear resulting states, we perform 2D DNS, with the numerical setup described in \S\ref{sec:dns_setup}. The centre mode produces only weak heat-transfer enhancement (HTE), whereas the convective mode yields substantial HTE and leads to two distinct nonlinear states, namely travelling waves and periodic orbits, accompanied by a hook-shaped polymer stress $\mathrm{tr}(\boldsymbol{\tau})$ (\S\ref{sec:six_cases}). Owing to its much stronger HTE, the convective mode is the focus of the subsequent analysis. A survey of the $Re$--$W$ parameter space identifies significant HTE at sufficiently large $Re$ and a certain range of $W$ (\S\ref{sec:parametric_re_w}). Analysis of the flow field links this behaviour to a wall-attached hook-shaped structure (\S\ref{sec:flow_field}). Finally, a perturbation kinetic energy budget confirms the elasto-buoyant nature of both travelling waves and periodic orbits, and shows that their sustainment results from a competition between buoyancy and polymer stresses (\S\ref{sec:energy_budget}).

%
%
\subsection{Problem setup}
\label{sec:dns_setup}

2D DNS is performed using the open-source pseudospectral code Dedalus \citep{burns_dedalus_2020} on the evolution equations for the nonlinear perturbation away from the base state:
\begin{equation}
\nabla \cdot {\boldsymbol{u'}} = 0,
\label{eq:nonlinear_perturbation-first}
\end{equation}
\begin{equation}
\frac{{\partial {\boldsymbol{u'}}}}{{\partial t}} 
+ {\boldsymbol{u'}} \cdot \nabla {\boldsymbol{u'}} 
+ {\boldsymbol{u'}} \cdot \nabla {{\boldsymbol{U}}} 
+ {{\boldsymbol{U}}} \cdot \nabla {\boldsymbol{u'}} 
= - \nabla p' 
+ \frac{{1 - \beta }}{{Re}}\nabla \cdot {\boldsymbol{\tau '}} 
+ \frac{\beta }{{Re}}{\nabla ^2}{\boldsymbol{u'}} 
+ Ri \rho' \boldsymbol{\hat{z}},
\label{eq:nonlinear_perturbation-NS}
\end{equation}
\begin{equation}
\begin{split}
\frac{\partial \boldsymbol{\tau}'}{\partial t} 
&+ \boldsymbol{u}'\!\cdot\!\nabla \boldsymbol{\tau}' 
+ \boldsymbol{u}'\!\cdot\!\nabla \boldsymbol{T} 
+ \boldsymbol{U}\!\cdot\!\nabla \boldsymbol{\tau}' 
- \nabla {\boldsymbol{u}'}^{T}\!\cdot\!\boldsymbol{\tau}' 
- \nabla {\boldsymbol{u}'}^{T}\!\cdot\!\boldsymbol{T} 
- \nabla \boldsymbol{U}^{T}\!\cdot\!\boldsymbol{\tau}' \\
&\quad - \boldsymbol{\tau}'\!\cdot\!\nabla \boldsymbol{u}' 
- \boldsymbol{\tau}'\!\cdot\!\nabla \boldsymbol{U}
- \boldsymbol{T}\!\cdot\!\nabla \boldsymbol{u}' 
+ \frac{1}{W}\boldsymbol{\tau}' 
= \frac{1}{W}\bigl(\nabla \boldsymbol{u}' + \nabla {\boldsymbol{u}'}^{T}\bigr) 
+ \frac{1}{{Re Sc}}\,\nabla^2 \boldsymbol{\tau}',
\end{split}
\end{equation}
\begin{equation}
\frac{{\partial \rho '}}{{\partial t}} 
+ {\boldsymbol{u'}} \cdot \nabla \rho ' 
+ {\boldsymbol{u'}} \cdot \nabla R 
+ {{\boldsymbol{U}}} \cdot \nabla \rho ' 
= \frac{1}{{Pr Re}}{\nabla ^2}\rho',
\label{eq:nonlinear_perturbation-end}
\end{equation}
where $\boldsymbol{U}$, $\boldsymbol{T}$, and $R$ denote the base-state variables in \eqref{eq:basestate}. Here, we add  a small global diffusion term ($\nabla^2 \boldsymbol{\tau}'$) which acts on the possibly large perturbation away from this base state to stabilise the DNS \citep{sureshkumar_linear_1995, dubief_first_2022}. Specifically, we use a fixed high Schmidt number $Sc=500$ for the convective mode and 1000 for the centre mode.

Simulations are initialised using the eigenfunction of the fastest-growing mode from the linear analysis, scaled by a small amplitude to capture the exponential growth regime. 
The centre-mode simulations utilise a polymer-diffusion-modified eigenfunction at $Sc = 1000$, subject to the boundary conditions defined in equation~\eqref{eq:mingdong_bc}, while the convective-mode simulations are initialised with the non-diffusive eigenfunction (i.e. $Sc = \infty$). The DNS is performed in computational domains whose streamwise length equals one wavelength, i.e.
\begin{equation*}
(x,z) \in \bigl[0, 2\pi/k] \times  [-1,1],
\end{equation*}
where $k$ is the wavenumber of the locally fastest growing mode. In the DNS, since $Sc \neq \infty$ we need to apply boundary conditions on the polymer stress. Focusing on the convective and centre modes, we apply the boundary conditions proposed by \cite{dong_asymptotic-analysis-inspired_2025} to eliminate the polymer diffusive instability (PDI) \citep{beneitez_polymer_2023}. These are
\begin{equation}
\frac{\partial \tau_{11}'}{\partial z} = 0, \quad 
\tau_{ij}'=0,\quad (i,j)\neq(1,1)
\qquad \text{at } z = \pm 1.
\label{eq:mingdong_bc}
\end{equation}

To quantify the level of heat transfer, a normalised (by the conductive heat flux) Nusselt number $Nu$ is employed:
\begin{equation}
Nu
:=
\frac{1}{2L_x}
\int_{0}^{L_x}
\left[
\left.\frac{\partial \rho / \partial z}{\partial R / \partial z}\right|_{z=1}
+
\left.\frac{\partial \rho / \partial z}{\partial R / \partial z}\right|_{z=-1}
\right] \, dx.
\label{eq:nusselt_origi}
\end{equation}

Since $Nu$ is generally time-dependent albeit fluctuating around a mean for statistically steady states, we characterise the heat transfer of the resulting state using its time-average, $\langle Nu \rangle$, defined as:
\begin{equation}
    \langle Nu \rangle = \frac{1}{T_2 - T_1}\int_{T_1}^{T_2} Nu(t) \, \mathrm{d}t.
    \label{eq:nuz_avg}
\end{equation}

Each simulation is run until a steady or periodic state is reached, and the integration window $t \in [T_1, T_2]$ is chosen to fall entirely within this regime. The heat-transfer enhancement (HTE) is quantified by:
\[{\rm{HTE}} = \frac{{{{\left\langle {Nu} \right\rangle }_{W > 0}} - {{\left\langle {Nu} \right\rangle }_{W = 0}}}}{{{{\left\langle {Nu} \right\rangle }_{W = 0}}}} \times 100\%, \]
where the Nusselt numbers ${\left\langle {Nu} \right\rangle }$ all belong to the same $Re$. We use the typical resolution range from $128 \times 144$ to $1024 \times 1152$ to guarantee the numerical stability of each simulation and ensure grid independence.

%
%
\subsection{Temporal evolution of representative cases} 
\label{sec:six_cases}

\begin{table}
  \centering
  \begin{tabular}{cccccccc}
    \toprule
    Case & $Re$ & $W$ & $\beta$ & $Ri$ & $k$ & $\omega_i$ (Sc = $\infty$) & $\omega_i$ (Sc = $1000$) \\
    \midrule
    Convective mode A & 50  & 0   & 1    & $-0.8$ & 2.8480 & 0.1840 & \\
    Convective mode B & 50  & 10  & 0.98 & $-0.8$ & 2.8480 & 0.1728 & \\
    Convective mode C & 50  & 25 & 0.98 & $-0.8$ & 2.6561 & 0.1515 & \\
    Convective mode D & 800 & 25  & 0.8  & $-0.5$ & 7.5646 & 0.2676 & \\
    Centre mode E     & 50  & 250 & 0.98 & $-0.2$ & 1.1498 & 0.0022 & 0.0032\\
    Centre mode F     & 50  & 250 & 0.98 & $0.2$  & 1.0723 & $9.9626\times10^{-4}$ & 0.0019\\
    \bottomrule
  \end{tabular}
  \caption{Summary of the linear stability analysis results of the six representative cases for DNS from Figure~\ref{fig:lsa_50_800}. The growth rates $\omega_i$ at $Sc = 1000$ are calculated by applying boundary conditions~\eqref{eq:mingdong_bc}.}
  \label{tab:six_DNS_summary}
\end{table}

%
%
\begin{figure}
\centerline{\includegraphics[width=0.999999\linewidth]{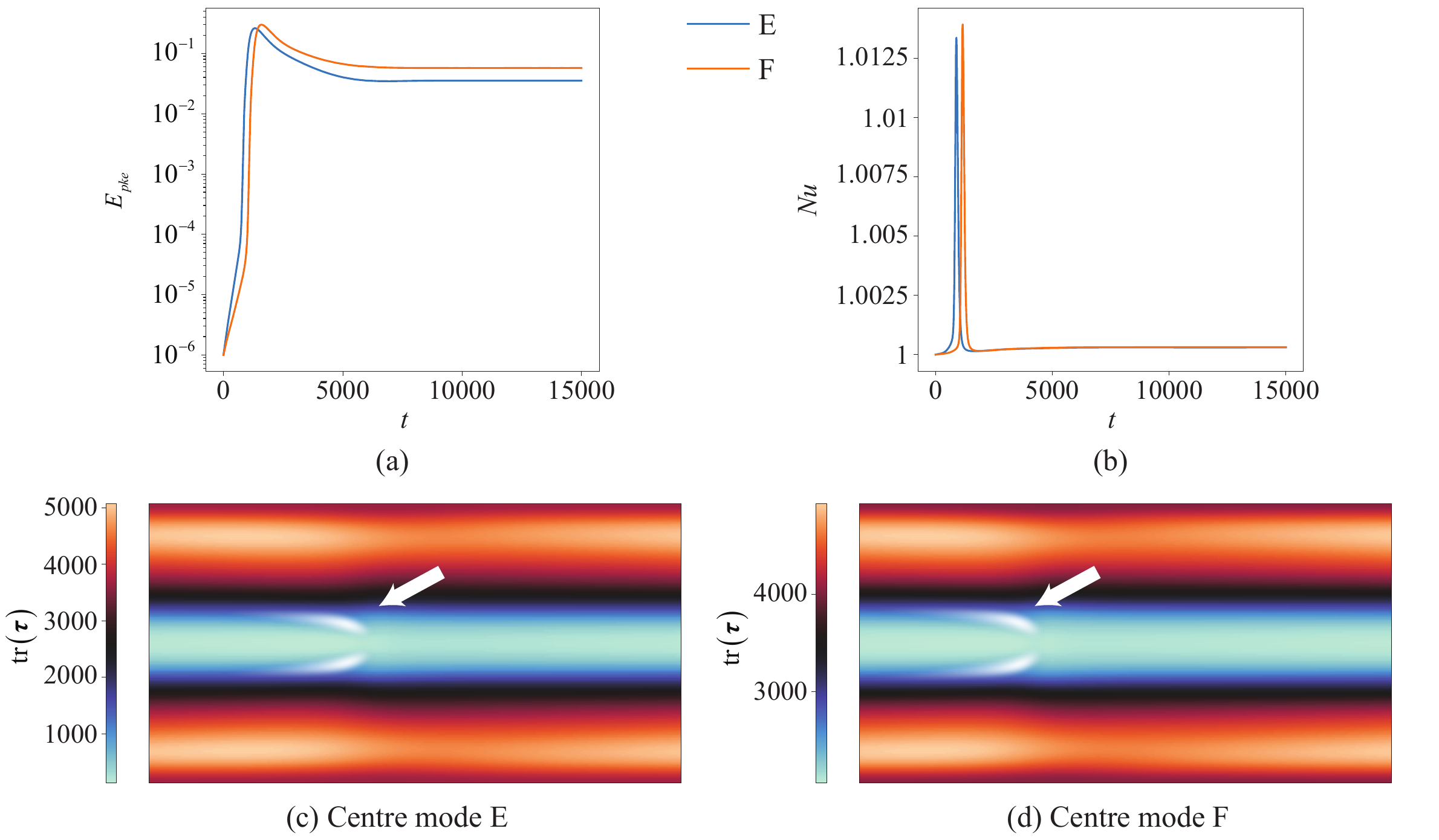}}
\caption{Time evolution of (a) perturbation kinetic energy $E_{pke}$ and (b) Nusselt number $Nu$ for centre modes E and F in table~\ref{tab:six_DNS_summary}, together with the resulting nonlinear travelling-wave structures in $\text{tr}(\boldsymbol{\tau})$ field. 
The growth rates of $E_{pke}$ for centre modes E and F are 0.0062 and 0.0038, respectively, which agree with $2\omega_i$ from the linear stability analysis (0.0064 and 0.0038). 
The $\operatorname{tr}(\boldsymbol{\tau})$ field is shown for (c) mode E and (d) mode F; the white shading overlays the fluctuation $\operatorname{tr}(\boldsymbol{\tau}) - \davg{\operatorname{tr}(\boldsymbol{\tau})}$, where $\davg{(\cdot)}$ is defined in equation~\eqref{eq:avg_operator}, marking the weak `arrowhead' at the centreline (white arrows).}

\label{fig:centre_cases}
\end{figure}

%
%
\begin{figure}
\centerline{\includegraphics[width=0.999999\linewidth]{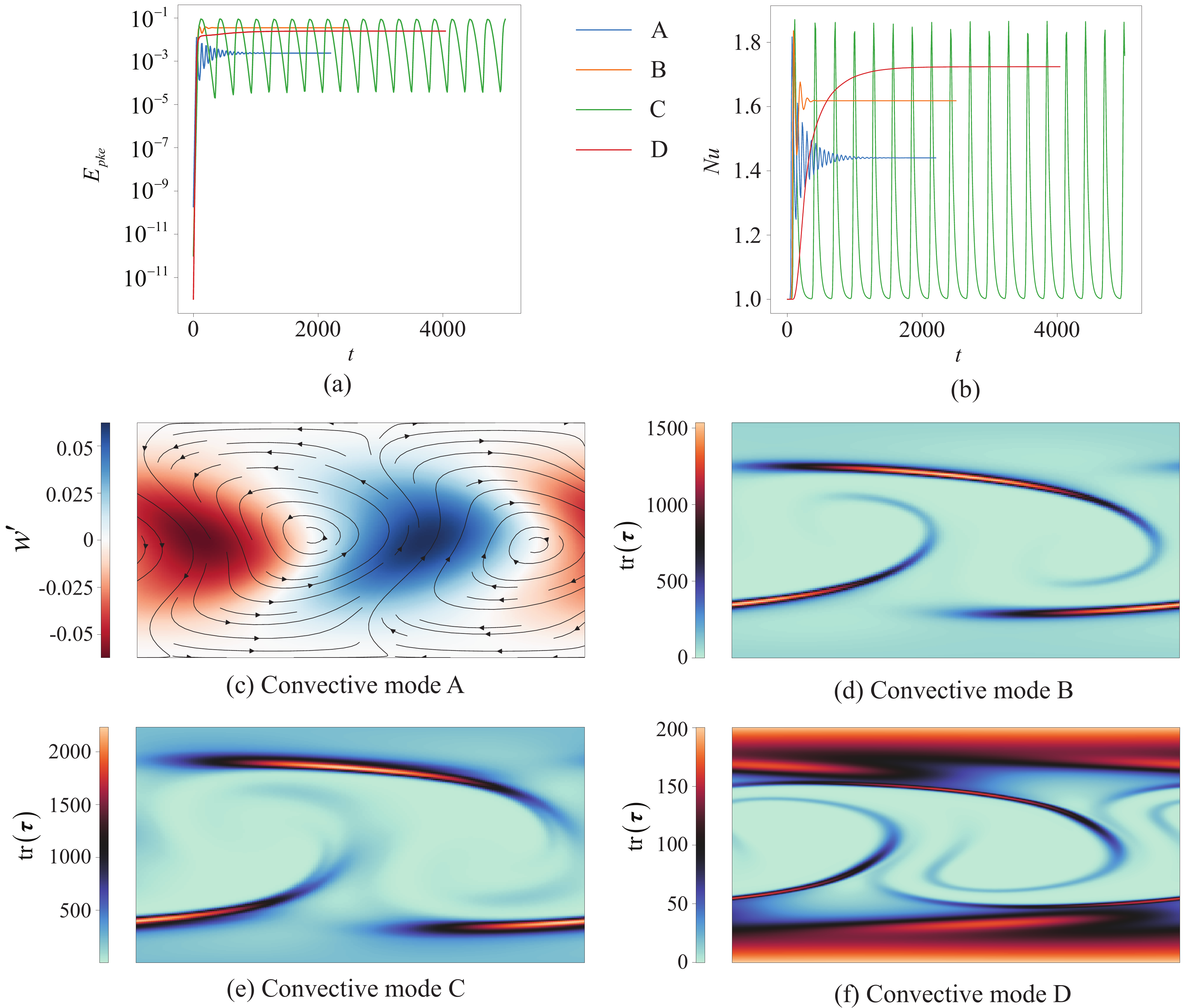}}
\caption{Time evolution of (a) perturbation kinetic energy $E_{pke}$ and (b) Nusselt number $Nu$ for convective modes A, B, C and D in table~\ref{tab:six_DNS_summary}, together with the resulting nonlinear states.
Panel (c) shows the vertical perturbation velocity $w'$ for mode A, overlaid with perturbation velocity streamlines; panels (d), (e) and (f) show the $\operatorname{tr}(\boldsymbol{\tau})$ field for modes B, C and D, respectively.
The field for mode C is shown at the instant when $Nu$ reaches its maximum during the periodic orbit; those for modes A, B, and D correspond to travelling-wave states and are stationary in their respective co-moving frames.
The growth rates of $E_{pke}$ for convective modes A, B, C, and D are 0.3678, 0.3456, 0.3026, and 0.5352, respectively, which agree with $2\omega_i$ from the linear stability analysis (0.3680, 0.3456, 0.3030, and 0.5352).}
\label{fig:convective_cases}
\end{figure}

In this section, we focus on the temporal evolution of the six representative cases, summarised in Table~\ref{tab:six_DNS_summary}, to characterise the nonlinear states and their associated heat transfer.
Here, the Newtonian convective mode~A acts as the baseline for the viscoelastic convective modes~B and~C. Conversely, owing to the computational cost of DNS at higher $Re$, only one convective mode~D at $Re=800$ is represented.

To compare with the linear modes, we show the time series of the perturbation kinetic energy
\[
    E_{pke} = \int_V \frac{1}{2} \left\| \boldsymbol{u}' \right\|^2 \, dV,
\]
of each case in Figure~\ref{fig:centre_cases}(a) and \ref{fig:convective_cases}(a), where $V$ denotes the computational domain. In all cases, $E_{pke}$ exhibits a clear initial phase of exponential growth, with the growth rates in agreement with linear stability analysis. This confirms that the inclusion of the diffusion term $\nabla^2\boldsymbol{\tau}$ does not significantly alter the DNS results from the linear stability predictions for the convective mode. Likewise, the centre-mode DNS are in exact agreement with the linear stability analysis at $Sc = 1000$.

For the centre mode, both modes E and F evolve into a travelling wave solution with steady $Nu$ (Figure~\ref{fig:centre_cases}(b)) and a nonlinear arrowhead structure \citep{Page_2020} in the $\operatorname{tr}(\boldsymbol{\tau})$ field (Figure~\ref{fig:centre_cases}(c,d)), characterised by a distinct V-shaped spatial organisation localised near the channel centreline (marked by the arrow). 
However, the associated vertical velocity perturbation $w'$ is too weak to convect heat efficiently across the channel, leading to a marginal increase of approximately $0.030\%$ in $\langle Nu\rangle$ relative to the laminar Newtonian counterpart. As a result, the centre mode~E and F cases and the corresponding arrowhead structure are ineffective for HTE in the present flow (as also suggested recently by \cite{LewyK25}).

In the Newtonian regime, the convective mode (Figure~\ref{fig:convective_cases}(c)) evolves into a transverse-roll structure. In the viscoelastic regime (Figure~\ref{fig:convective_cases}(d-f)), elastic forces modify the Newtonian convective mode giving rise to two classes of resulting states:
\emph{elasto-buoyant travelling waves} (modes B and D) and \emph{elasto-buoyant periodic orbits} (mode C). Correspondingly, the time series of $Nu$ (Figure~\ref{fig:convective_cases}(b)) is either steady for the travelling waves or oscillates about its mean value for the periodic orbits (the elasto-buoyant nature of these states will be discussed in \S\ref{sec:energy_budget}).
The \emph{hook-like structure} in the $\operatorname{tr}(\boldsymbol{\tau})$ field is observed in both travelling wave (Figure~\ref{fig:convective_cases}(d,f)) and the moment when $Nu$ attains its peak in the cycle period of periodic states (Figure~\ref{fig:convective_cases}(e)). This structure is characterised by two high-stress curves (large $\operatorname{tr}(\boldsymbol{\tau})$ regions) on opposite sides of the channel centreline offset in the streamwise direction.
Moreover, the periodic orbit (mode~C) can undergo a relaminarisation cycle when $W$ exceeds a critical value. Specifically, the perturbation first experiences a phase of linear growth, during which $Nu$ rises to a temporal maximum, and then decays until $Nu$ approaches unity. At this near-relaminarised state, the perturbation field assumes an unstable eigenfunction-like structure, from which a new phase of linear growth emerges, and the process repeats periodically. The mechanism of the relaminarisation cycle will be further discussed in \S\ref{sec:energy_budget}.
In terms of the heat transfer efficiency, convective modes B ($Re = 50$, $\langle Nu \rangle = 1.618$) and D ($Re = 800$, $\langle Nu \rangle = 1.724$) attain larger values than their respective Newtonian counterparts ($\langle Nu \rangle = 1.440$ (mode~A) and $1.210$). In contrast, convective mode C yields a lower $\langle Nu \rangle$ ($1.004$) compared to its Newtonian equivalent (mode A).

Since only very minor HTE is observed in centre mode, we will henceforth focus on the convective modes that can enhance the heat-transfer at both $Re = 50$ (mode B) and $Re = 800$ (mode D). 

%
%
\subsection{Regime diagram}
\label{sec:parametric_re_w}

%
%
\begin{figure}
\centering
\begin{subfigure}{0.49\textwidth}
    \centering
    \includegraphics[width=\textwidth]{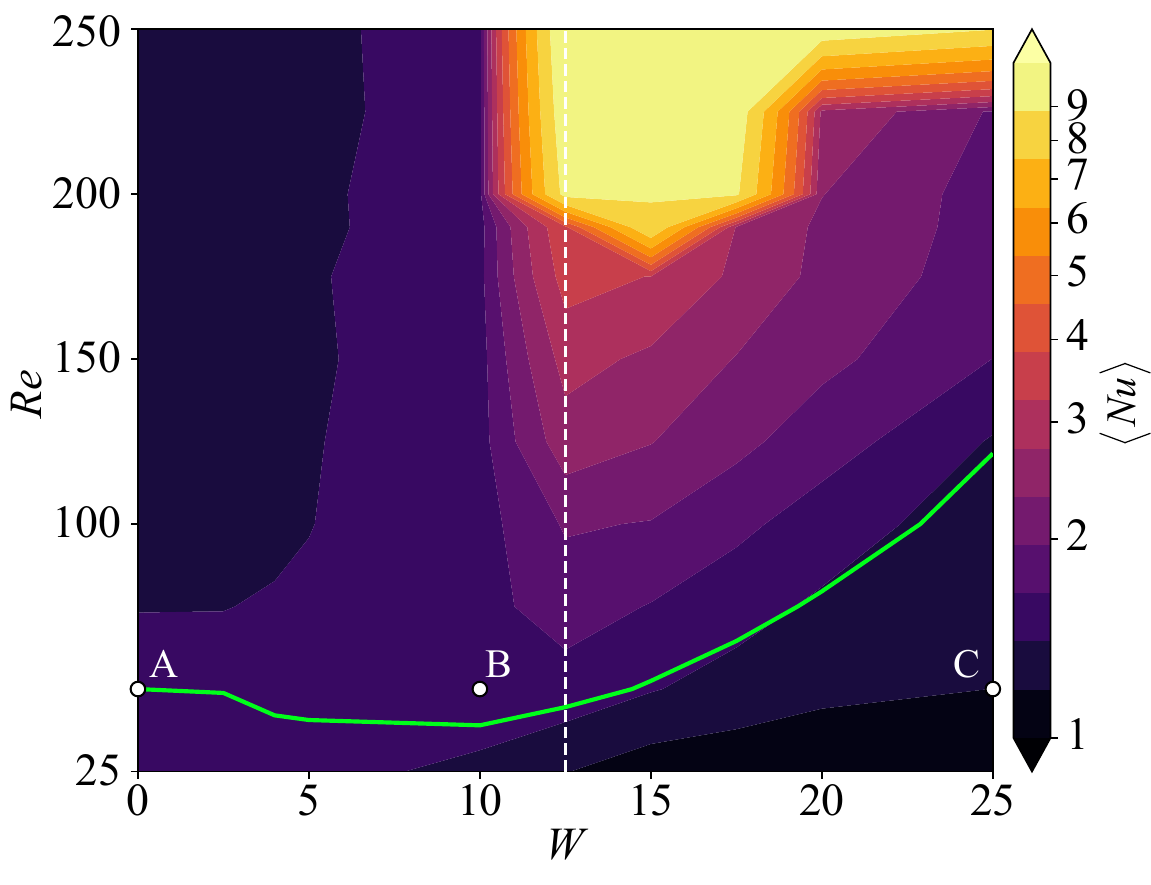}
    \caption{}
\end{subfigure}
\hfill
\begin{subfigure}{0.46\textwidth}
    \centering
     \includegraphics[width=\textwidth]{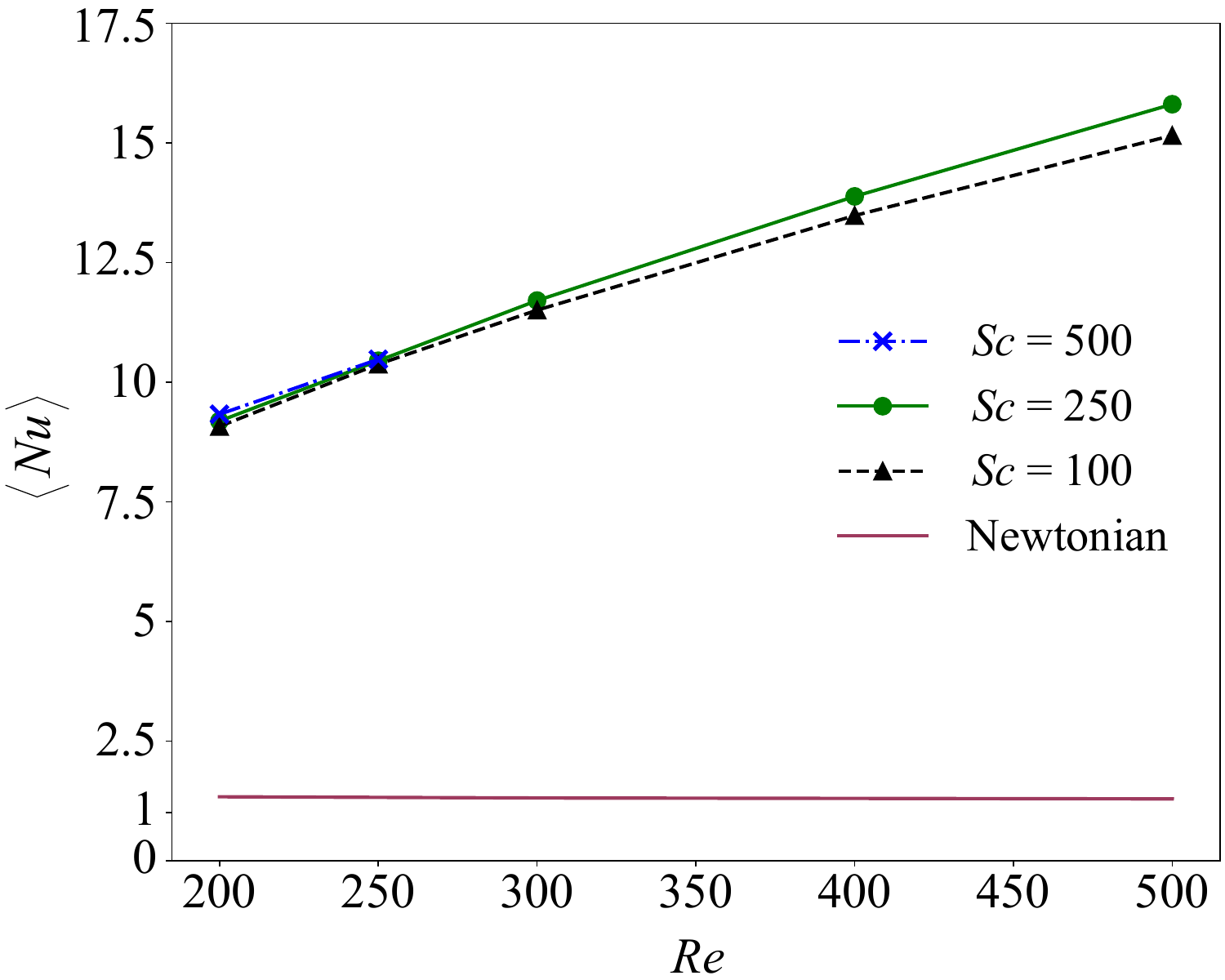}
    \caption{}
\end{subfigure}
  \caption{Variations of $\langle Nu\rangle$ with $Re$ and $W$. (a) Contours of $\langle Nu\rangle$ in the $Re$--$W$ parameter space for fixed $\beta = 0.98$ and $Ri = -0.8$. Marks A, B and C correspond to the convective modes A, B and C shown in Figure~\ref{fig:lsa_50_800}(a). The white dashed line marks $W=12.5$ and corresponds to the further exploration in panel (b). The green solid line indicates zero HTE. (b) Variations of $\langle Nu\rangle$ with $Re$ at different $Sc$ for fixed $\beta = 0.98, Ri = -0.8$ and $W = 12.5$. Red solid line indicates the Newtonian counterpart.}
\label{fig:nu_heatmap}
\end{figure}


A parametric study of $(Re,W)\in[25,250]\times[0,25]$ was conducted for the convective mode at fixed $\beta = 0.98$ and $Ri = -0.8$ to clarify the HTE behaviour, as shown in Figure \ref{fig:nu_heatmap} (mode D at $Re = 800$ proved too costly to simulate).
In Figure \ref{fig:nu_heatmap}(a), no HTE is observed for $Re \lesssim 37.5$. However, above this $Re$, larger $\langle Nu \rangle$ with a maximum within the range $W \in [10, 15]$, is obtained. Here, $\langle Nu \rangle$ increases monotonically with $Re$ and undergoes a sudden transition near $Re \approx 200$ to the upper yellow region with substantially high $\langle Nu \rangle$. A maximum $\langle Nu \rangle$ of 10.476 is achieved at the largest $Re = 250$ considered with $W = 12.5$. This is approximately 700\% more than its Newtonian counterpart ($\langle Nu \rangle = 1.323$). To the authors' knowledge, an increase of this magnitude has not been reported before.

In Figure~\ref{fig:nu_heatmap}(b), we run simulations at higher $Re$ with fixed $W = 12.5$ to extend beyond the peak HTE observed in Figure~\ref{fig:nu_heatmap}(a). Note that the results are insensitive to $Sc$.  In all these cases, the flows are travelling waves. The monotonic increase of $\langle Nu \rangle$ with rising $Re$ continues towards the highest $Re=500$.
Specifically, we observed a maximum $\langle Nu \rangle = 15.807$ at $Re=500$ and $Sc=250$, representing a significant HTE of approximately 1100\% compared to the Newtonian case ($\langle Nu \rangle = 1.291$). This HTE looks set to continue to yet larger $Re$.

%
%
\subsection{Mechanism behind enhanced heat transfer}
\label{sec:flow_field}

%
%
\begin{figure}
\centerline{\includegraphics[width=0.999\linewidth]{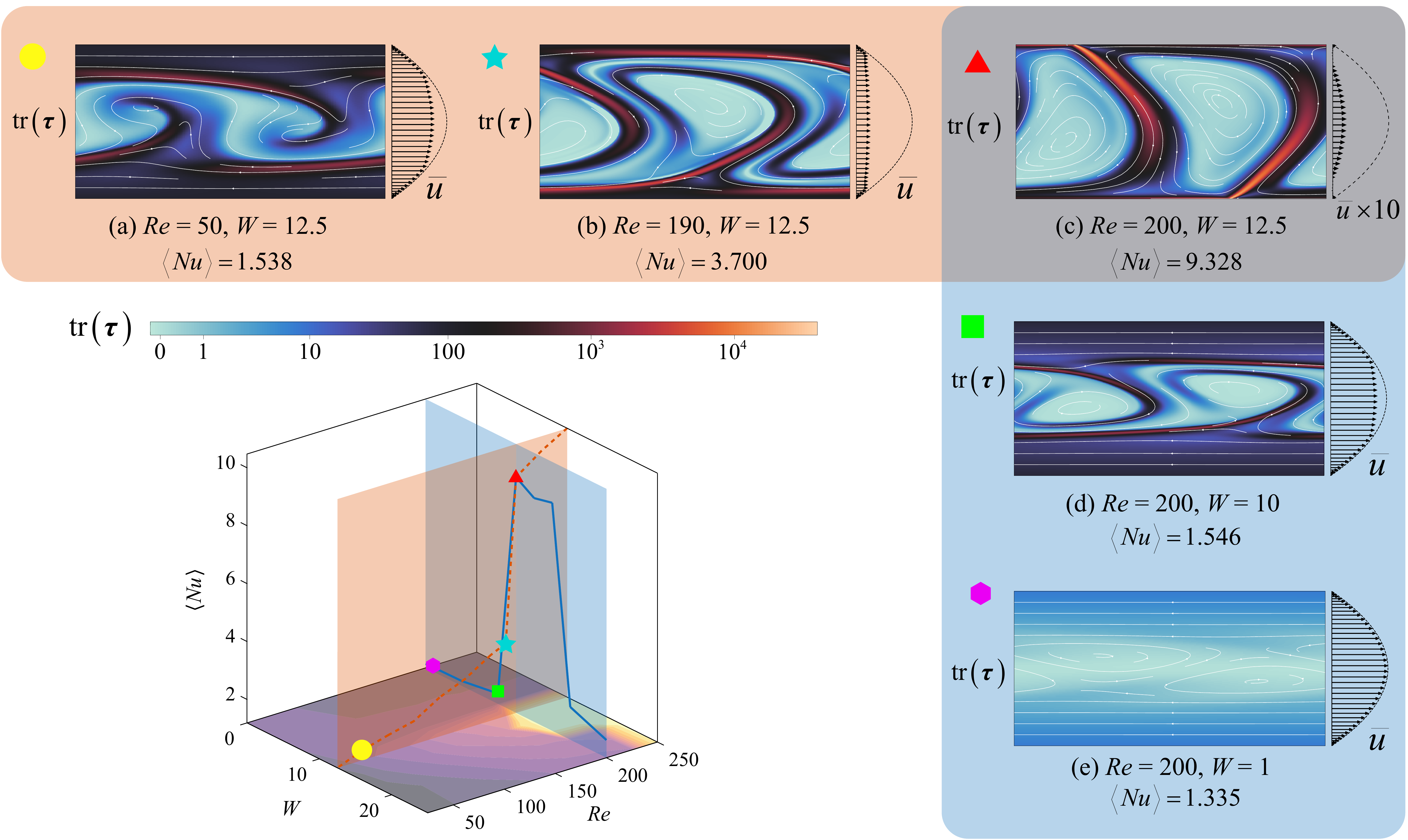}}
\caption{Flow structures for fixed $\beta = 0.98$ and $Ri = -0.8$, shown alongside $\langle Nu \rangle$ contours. Note that the streamlines are computed in a reference frame moving with the phase speed $c_p$, whereas the flow-rate profile $\overline{u}(z)$, marked by black arrows, is shown in the laboratory frame. The parabolic velocity profile, marked by black dashed line, corresponding to the base state, $U_0 = 1 - z^2$, is shown for reference. Panels (a)--(d) represent travelling-wave states that are stationary in their respective co-moving frames, whereas panel (e) shows a periodic orbit at the instant when $Nu$ attains its maximum.}

\label{fig:relative_flow_field}
\end{figure}

%
%
\begin{figure}
\centerline{\includegraphics[width=0.5\linewidth]{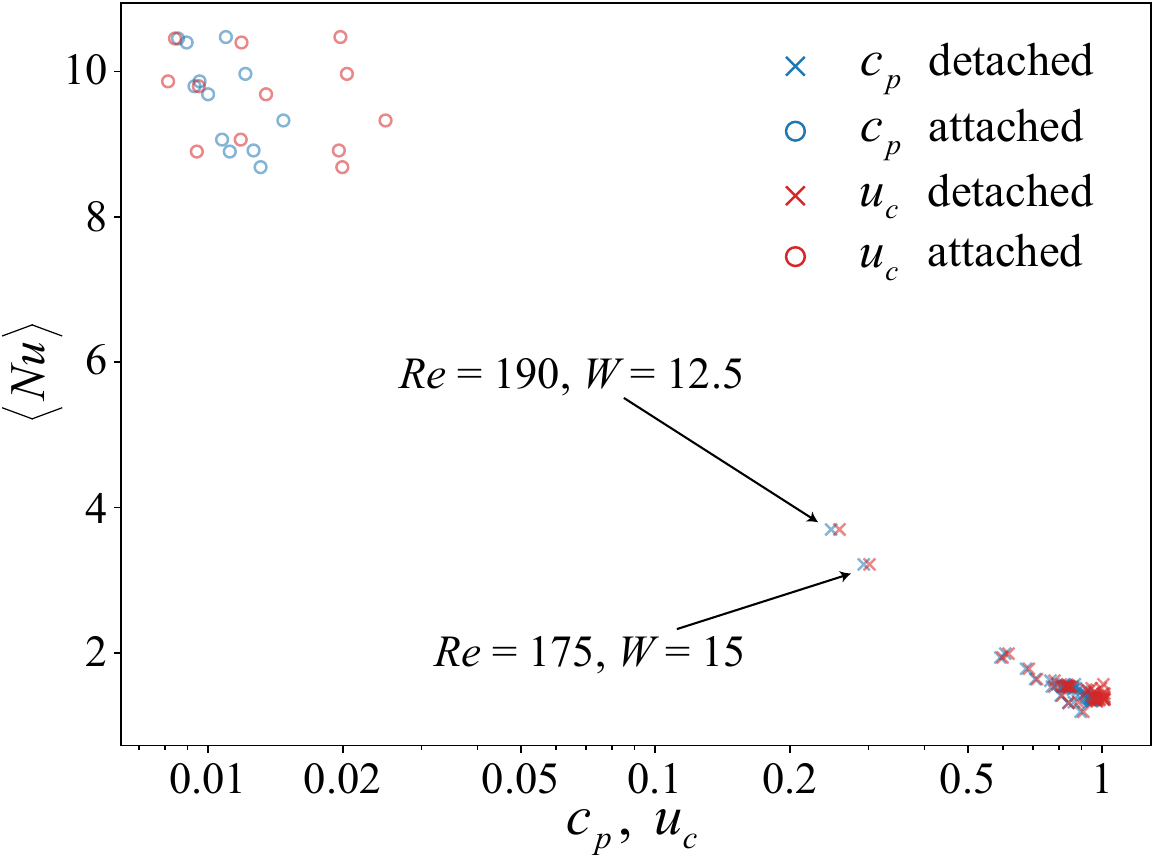}}
\caption{Time-averaged Nusselt number $\langle Nu \rangle$ against the phase speed $c_p$ (blue) and the centreline velocity $u_c$ (red) for the 60 travelling-wave cases in Figure~\ref{fig:nu_heatmap}(a). Crosses denote wall-detached hooks and circles wall-attached hooks.}
  \label{fig:nu_vs_cp_uc}
\end{figure}

%
%
\begin{figure}
\centerline{\includegraphics[width=0.999\linewidth]{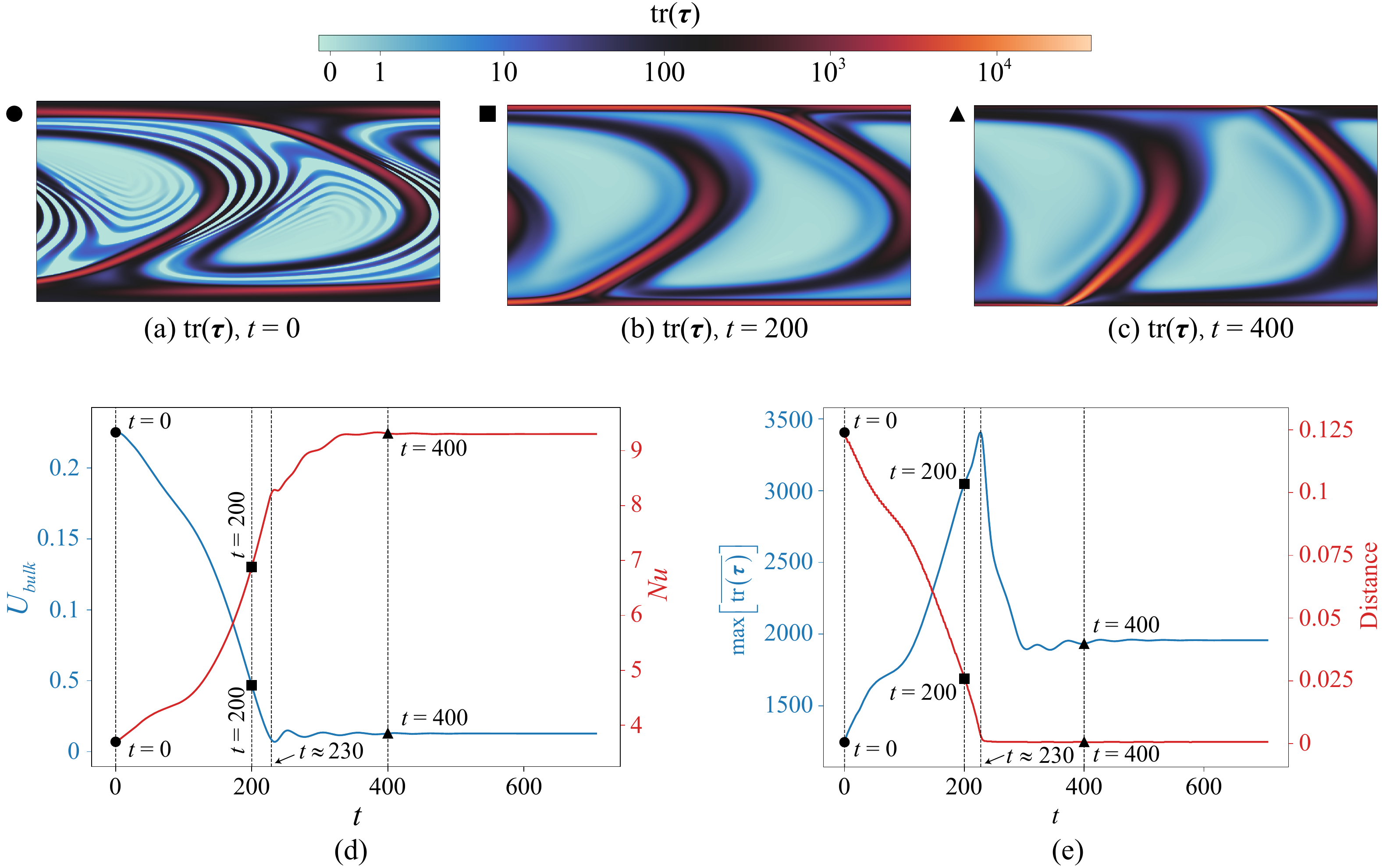}}
\caption{Transition from wall-detached to wall-attached hook structures at $Re=200, W = 12.5, \beta = 0.98$ and $Ri = -0.8$: (a-c) instantaneous fields of $\operatorname{tr}(\boldsymbol{\tau})$ at $t = 0, 200, 400$ and (d) time evolution of bulk velocity $U_{bulk}$ and Nusselt number $Nu$, (e) time evolution of $\max\!\left[\overline{\operatorname{tr}(\boldsymbol{\tau})}\right]$ and its wall-normal distance.}
  \label{fig:tr_evolution}
\end{figure}

Motivated by the significant HTE of the convective-mode nonlinear states in the preceding section, we now examine the flow field to understand the mechanism through which elasticity promotes heat transfer.
We compute the velocity fields
\[\boldsymbol{u}_{rel} = {\boldsymbol{u}} - {c_p}{\boldsymbol{\hat x}},\]
in a frame moving with the phase speed $c_p$ for several travelling wave cases. 
Figure~\ref{fig:relative_flow_field} shows that the rise in $\langle Nu\rangle$ is accompanied by a progressive strengthening of the hook structure. In each case the hook acts as an effective `polymer wall', with the streamlines in the co-moving frame following the high-stress sheets rather than crossing them.

In Figure~\ref{fig:relative_flow_field}(a,b,d,e), hook structures are observed which are not attached to the wall. For these cases, the flow-rate profiles
$\overline{u}(z)$ (where the overline denotes averaging in the $x$ direction at the selected instant) are reduced relative to the base state profile $U_0$.
In panel (e), this reduction is weak due to the small $W$.
The detached hooks act as moving walls in the channel interior, deflecting the fluid vertically around them and reducing the flow rate.
From Figure~\ref{fig:relative_flow_field}(a) to (b) or (e) to (d), the hook structures progressively migrate towards the channel walls with increasing $Re$ or $W$. This outward migration and the increase in their magnitude are accompanied by an increase in the Nusselt number $\langle Nu \rangle$.

At sufficiently large $Re$ and a certain range of $W$
(Figure~\ref{fig:relative_flow_field}(c)), the hook structure attaches to the wall. 
The corresponding Nusselt number $\langle Nu \rangle$ experiences a sharp increase, leading to the yellow region in Figure~\ref{fig:nu_heatmap}(a).
This boundary interaction alters the global flow topology by driving a strong flow from the wall toward the channel centre, forcing the fluid into distinct pairs of strong counter-rotating rolls.
These rolls drastically enhance the vertical transport of heat, triggering a massive surge in the heat transfer but also strongly reducing the flow rate.
This wall-attached hook is independent of the boundary condition and is also observed in Neumann and diffusion-free \citep{sureshkumar_linear_1995,lin_weakening_2025} boundary conditions for polymer stresses. A similar `sheet-like' structure in viscoelastic Rayleigh-Bénard convection flow has been  reported previously by \cite{dubief_heat_2020}.

The two families of hooks are further distinguished by how fast they travel. Figure~\ref{fig:nu_vs_cp_uc} shows $\langle Nu \rangle$ against the phase speed $c_p$ and the centreline velocity $u_c$ for all 60 travelling-wave cases of Figure~\ref{fig:nu_heatmap}(a). Both speeds show a clear inverse correlation with $\langle Nu \rangle$: the more slowly the hook travels, the larger the heat transfer. The detached hooks travel with the flow, at $96$--$99\%$ of the centreline velocity ($c_p \approx 0.25$--$0.96$ in Figure~\ref{fig:nu_heatmap}(a), so the two sets of crosses nearly coincide), and their heat transfer is modest. The wall-attached hooks travel much more slowly ($c_p \approx 0.01$), with the centreline velocity itself reduced to $O(10^{-2})$, and attain a significant heat transfer. Two intermediate cases ($Re = 190$, $W = 12.5$ and $Re = 175$, $W = 15$) lie between the two clusters, consistent with the smooth evolution from the detached to the attached structure discussed below.

In general, HTE strategies typically incur a larger pressure drop and therefore require greater pumping power \citep{manglik_heat_2003}. The reduced flow rate $\overline{u}$ associated with the HTE regime shown in Figure~\ref{fig:relative_flow_field}(c) in particular indicates that our system is no exception. This trade-off is examined further in \S\ref{sec:engineering}.

To investigate the transition between wall-detached and wall-attached hook structures, we select a snapshot at $Re = 190$ and $W = 12.5$ (a travelling wave) and use it as the initial condition for a simulation at an increased Reynolds number of  $Re = 200$ in Figure~\ref{fig:tr_evolution}.
As shown in Figure~\ref{fig:tr_evolution}(a--c), the wall-detached hook progressively evolves into a wall-attached hook as time progresses. During this transition, the Nusselt number $Nu$, as shown in Figure~\ref{fig:tr_evolution}(d), increases, whereas the bulk velocity $U_{bulk}$ decreases; both quantities subsequently approach a plateau where the wall-attached hook is realised. 
To characterise the emergence of the wall-attached structure, we track both the maximum of the streamwise-averaged stress trace $\overline{\operatorname{tr}(\boldsymbol{\tau})}$ and the wall-normal location at which this maximum occurs. As shown in Figure~\ref{fig:tr_evolution}(e), the maximum of $\overline{\operatorname{tr}(\boldsymbol{\tau})}$ increases monotonically and reaches a peak at approximately $t \approx 230$, at which time $U_{{bulk}}$ also attains a transient minimum (Figure~\ref{fig:tr_evolution}(d)). Meanwhile, the corresponding wall-normal distance decreases monotonically and becomes effectively zero at the same instant. Thereafter, it remains nearly unchanged. We therefore identify this instant as the onset of the wall-attached structure. Subsequently, the maximum value of $\overline{\operatorname{tr}(\boldsymbol{\tau})}$ decreases to approximately $2000$ and then remains approximately constant.
Since no abrupt change is observed during this transition, this smooth evolution further suggests that the wall-detached and wall-attached structures are not distinct states, but rather two different manifestations of the same underlying dynamical state.

%
%
\subsection{Energy analysis}
\label{sec:energy_budget}

%
%
\begin{figure}
  \centerline{\includegraphics[width=0.65\linewidth]{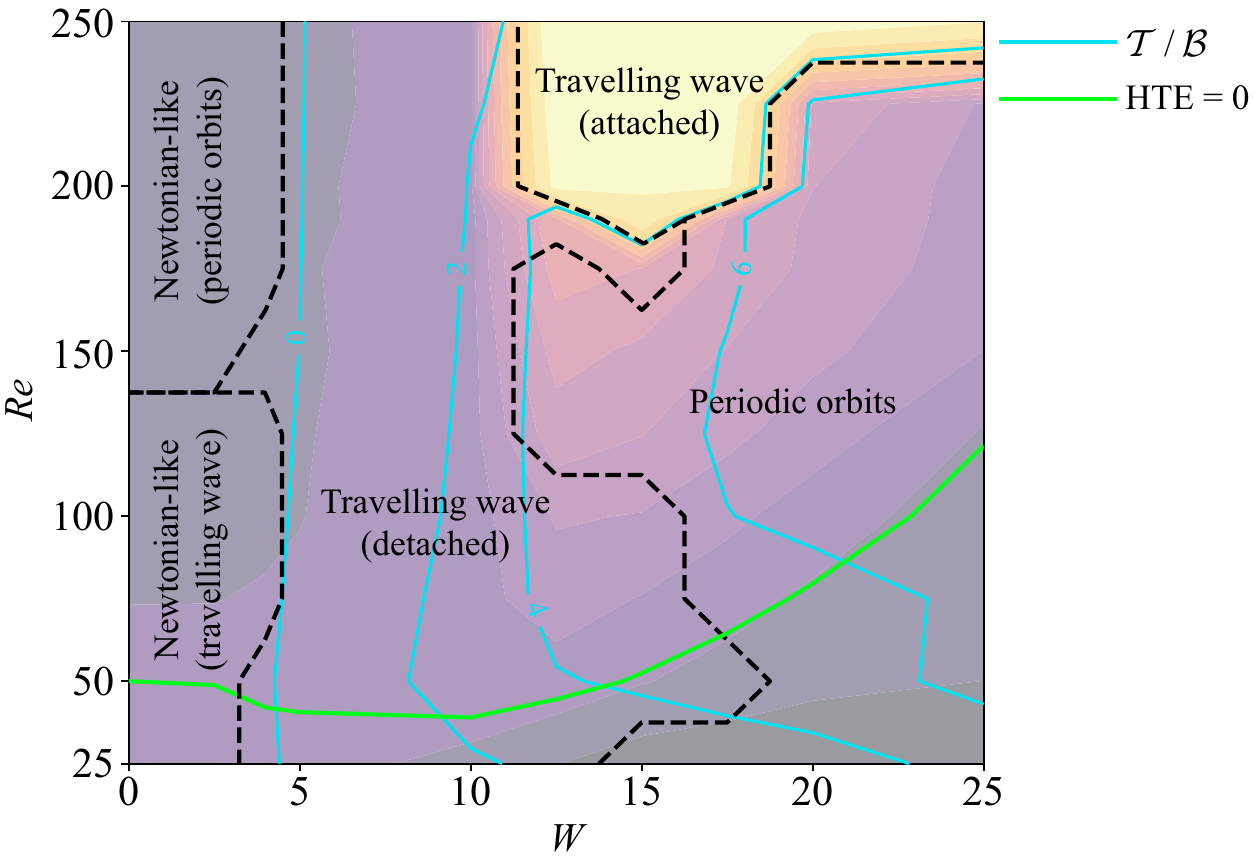}}
  \caption{Contour of energy budget ratio $\mathcal{T}/\mathcal{B}$ and the flow regime classification overlapped with contours of $\langle Nu\rangle$ corresponding to Figure~\ref{fig:nu_heatmap}(a). The black dashed line represents the boundary between different flow regimes.}
  \label{fig:heatmapPKE}
\end{figure}

In this section, we perform a perturbation kinetic energy (PKE) budget analysis, based on perturbations about the laminar base state, to investigate the mechanisms sustaining the nonlinear states.
The PKE budget isolates the contributions of processes that destabilise the flow relative to the laminar base state. This perspective is particularly relevant here because most of the states we study are travelling waves or periodic orbits which, unlike fully developed turbulence that is sustained by self-generated fluctuations around a turbulent mean, are maintained primarily through a modification of the base state. It also links to the Nusselt number in \eqref{eq:nusselt_origi}, which measures the ratio of convective heat transfer in the nonlinear state to conductive heat transfer in the laminar base state. By contrast, the typical turbulent kinetic energy budget, formulated in terms of fluctuations about the statistical mean state \citep{pope_turbulent_2001}, is primarily less relevant here (see Appendix~\ref{app:fke_budget}). 

We derive the PKE budget equation from the nonlinear perturbation Navier-Stokes equations given in \eqref{eq:nonlinear_perturbation-NS} by taking the inner product with $\boldsymbol{u'}$ and rearranging the terms (e.g. see \cite{zhang_linear_2013} although we extend their analysis beyond the linear growth regime). Applying the spatiotemporal averaging operator $\davg{(\cdot)}$, defined as
\begin{equation}
    \left[\kern-0.15em\left[ (\cdot) \right]\kern-0.15em\right]_z = \frac{1}{T_2 - T_1}\int_{T_1}^{T_2} \left[ \frac{1}{L_x}\int_0^{L_x} (\cdot) \, \mathrm{d}x \right] \mathrm{d}t,
\label{eq:avg_operator}
\end{equation}
where the integration window matches the one used for the averaged Nusselt number $\langle Nu \rangle$ in equation~\eqref{eq:nuz_avg}, yields the final PKE budget equation:
\begin{equation}
    \frac{\partial k}{\partial t} + \partial_j \mathcal{F}_j
    = \mathcal{P} + \mathcal{B} + \mathcal{T} + \mathcal{D},
    \label{eq:PKE_equation}
\end{equation}
where terms are defined as follows:

\begin{itemize}
    \item Kinetic energy. $k = \frac{1}{2}\davg{{u_i}'{u_i}'}$,
    \item Transport flux. ${\cal F}_j = \davg{p'{u_j}^\prime}  - \frac{{1 - \beta }}{{Re}}\davg{{u_i}^\prime {\tau _{ij}}^\prime}  - \frac{\beta }{{Re}}{\partial _j}{k}  + \tfrac{1}{2}\davg{{u_i}^\prime{u_i}^\prime{u_j}^\prime} + {U_j}{k}$,
    \item Production. $\mathcal{P} = - \davg{{u_i}^\prime {u_j}^\prime}{\partial _j}{U_i}$,
    \item Buoyancy flux. $\mathcal{B} = Ri\davg{{u_z}^\prime\rho'} $,
    \item Polymer conversion. $\mathcal{T} =  - \frac{{1 - \beta }}{{Re}}\davg{{\tau _{ij}}^\prime {\partial _j}{u_i}^\prime} $,
    \item Viscous conversion. $\mathcal{D} = - \frac{\beta }{{Re}}\davg{{\partial _j}{u_i}^\prime {\partial _j}{u_i}^\prime}$.
\end{itemize}

Since the production term $\mathcal{P}$ remains consistently negative and the buoyancy flux term $\mathcal{B}$ remains consistently positive across the entire $Re$--$W$ parameter space (Figure~\ref{fig:nu_heatmap} (a,b)), the distinction between buoyancy-driven and elasto-buoyant regimes is governed essentially by the sign of $\mathcal{T}$. Consequently, we utilise the ratio $\mathcal{T}/\mathcal{B}$ to classify the regime and quantify the relative importance of elastic forces compared to buoyancy forces.

In Figure~\ref{fig:heatmapPKE}, the $Re$--$W$ parameter space is partitioned into five distinct regions.
At small $W\lesssim5$, the flow is essentially Newtonian-like: its dynamical and statistical behaviour closely matches that of the corresponding Newtonian case at the same Reynolds number. Within this regime, we further identify two subregions depending on $Re$: a Newtonian-like periodic-orbit regime for $Re \gtrsim 140$, and a Newtonian-like travelling-wave regime at lower $Re$. The HTE of these cases is less than $1\%$. $\mathcal{T}/\mathcal{B}<0$, suggesting that polymers are extracting energy from the PKE reservoir.
The flow structures remain largely unchanged from their Newtonian counterparts, apart from the appearance of a weak hook structure in the $\operatorname{tr}(\boldsymbol{\tau})$ field. 

As $W$ exceeds the pre-onset threshold, elastic effects become non-negligible and travelling waves with distinct detached hooks emerge, accompanied by enhanced heat transfer. Moreover, for $Re \gtrsim 200$, these detached waves transition to attached travelling waves with a marked increase in HTE. In this regime, $\mathcal{T}/\mathcal{B}>0$, indicating that polymer stresses contribute to sustaining the travelling solutions; the resulting states are therefore elasto-buoyant in nature.

On the other hand, for relatively small $Re \lesssim 200$, the elasto-buoyant travelling wave can transition to an elasto-buoyant periodic-orbit regime, in which $\mathcal{T}/\mathcal{B}$ is predominantly positive. Substantial HTE is also achieved, with a maximum typically attained for $W \in [10,15]$. Note that in the periodic-orbit regime the hook structures remain detached from the wall, whereas they attach to the wall only in the travelling-wave solutions.

\begin{figure}
\centerline{\includegraphics[width=0.7\linewidth]{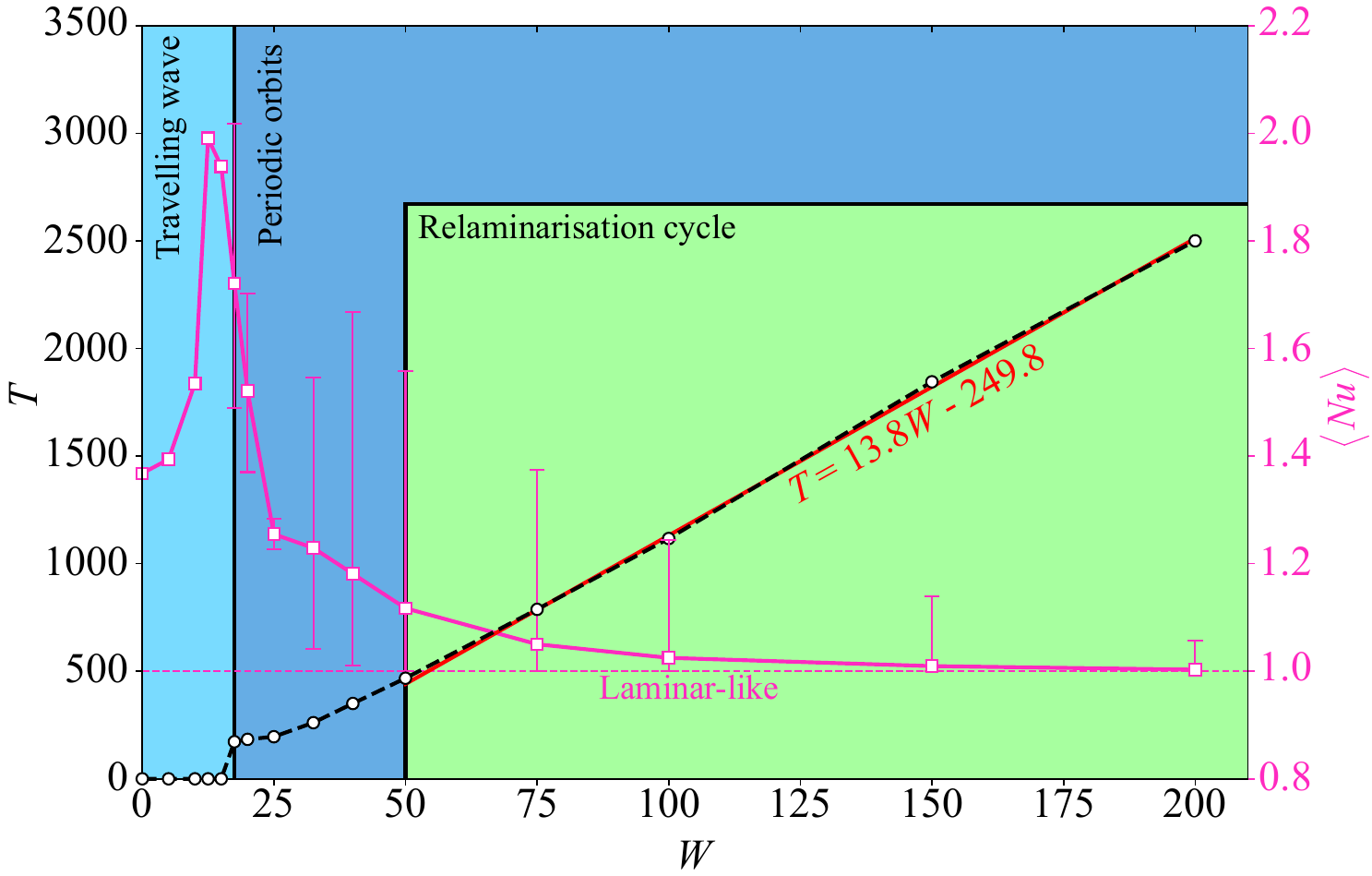}}
\caption{Cycle period $T$ and time-averaged Nusselt number $\langle Nu\rangle$ as a function of $W$ at $Re=100$ for fixed $\beta = 0.98$ and $Ri = -0.8$. The black dashed line represents $T$. The pink solid line represents $\langle Nu \rangle$, with error bars indicating the full range (maximum to minimum) of $Nu$ in the periodic orbits, while the pink dashed line denotes the baseline $\langle Nu \rangle=1$. The red solid line corresponds to the linear fit of $T$ versus $W$ when $W\ge50$, with $T = 13.8W - 249.8$.}
  \label{fig:period_varying}
\end{figure}

A competition between polymer and buoyancy forces in sustaining the nonlinear state can also be inferred from Figure~\ref{fig:heatmapPKE}.
After $W$ exceeds a critical value, the resulting state transitions from a travelling wave to a periodic orbit. As $W$ increases further, the magnitude of HTE decreases monotonically and eventually gives way to heat-transfer suppression. This correlation with $W$ suggests that the emergence of periodic orbits is driven by elastic effects. 
This interpretation is further supported by the cycle period $T$ in Figure~\ref{fig:period_varying}. The resulting state becomes a periodic orbit at $W \approx 17.5$. As $W$ increases, the period $T$ increases monotonically, while the minimum value of $Nu$ over each cycle decreases. For $W \ge 50$, the flow undergoes a relaminarisation cycle, characterised by the periodic transient approach of the minimum $Nu$ to nearly unity.
Correspondingly, at the minimum $Nu$ instant, the flow structure in the $\operatorname{tr}(\boldsymbol{\tau})$ field evolves from a weak hook-like structure at lower $W$ to an unstable eigenfunction-like structure at larger $W$ in the relaminarisation cycle regime.
In this regime, the flow undergoes repeated cycles of linear growth, decay, and renewed linear growth. The cycle period $T$ increases approximately linearly with $W$ (or, equivalently, with the relaxation time), indicating that $T$ is closely tied to the processes of polymer stretching and relaxation. As reflected by the ratio contours in Figure~\ref{fig:heatmapPKE}, the monotonic increase of $\mathcal{T}/\mathcal{B}$ with rising $W$ signifies a growing competition between these two forces. Polymer stresses become increasingly dominant as $W$ rises and once $W$ exceeds a critical threshold, these stresses eventually surpass the buoyancy effect. The observed linear scaling serves as a signature of this transition.

%
%

%
%
\section{Engineering implications}
\label{sec:engineering}

%
%
\begin{figure}
\centering
\begin{subfigure}{0.49\textwidth}
    \centering
    \includegraphics[width=\textwidth]{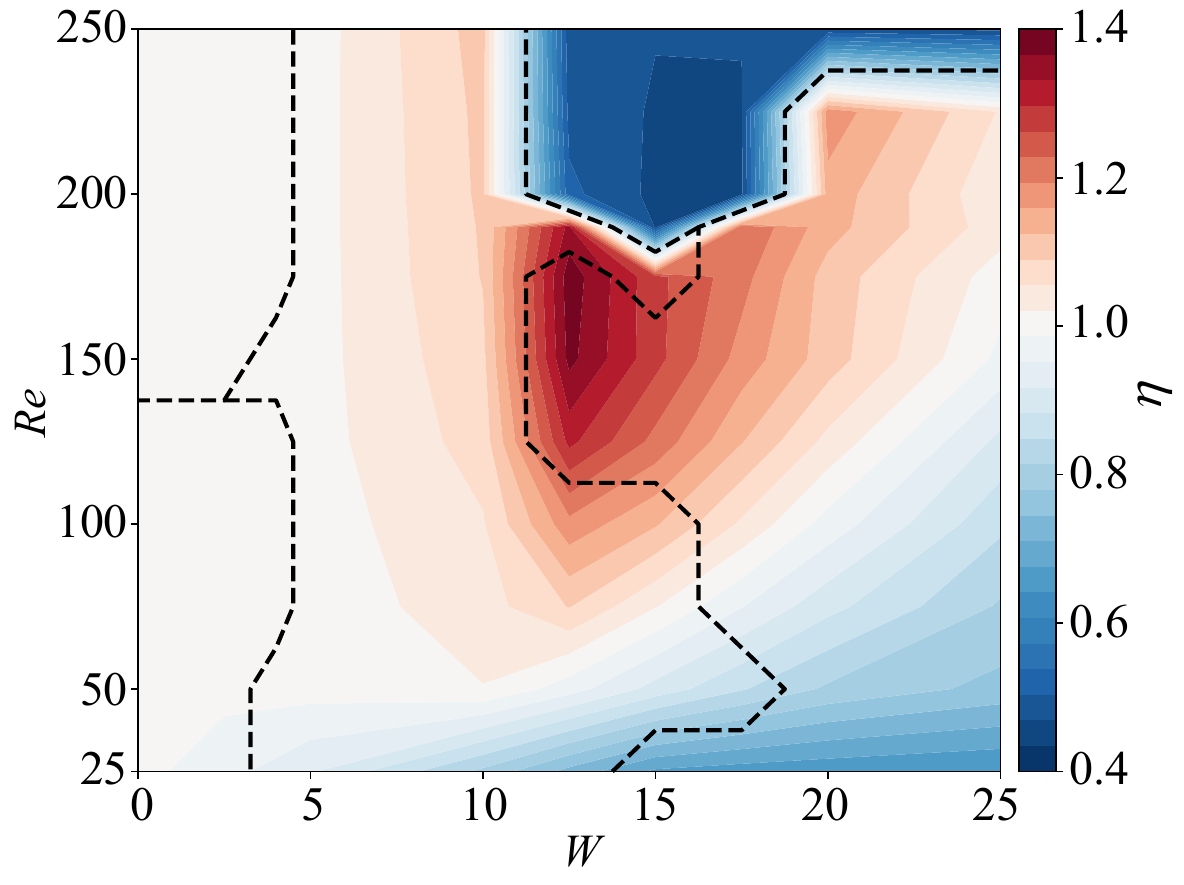}
    \caption{}
\end{subfigure}
\hfill
\begin{subfigure}{0.49\textwidth}
    \centering
     \includegraphics[width=\textwidth]{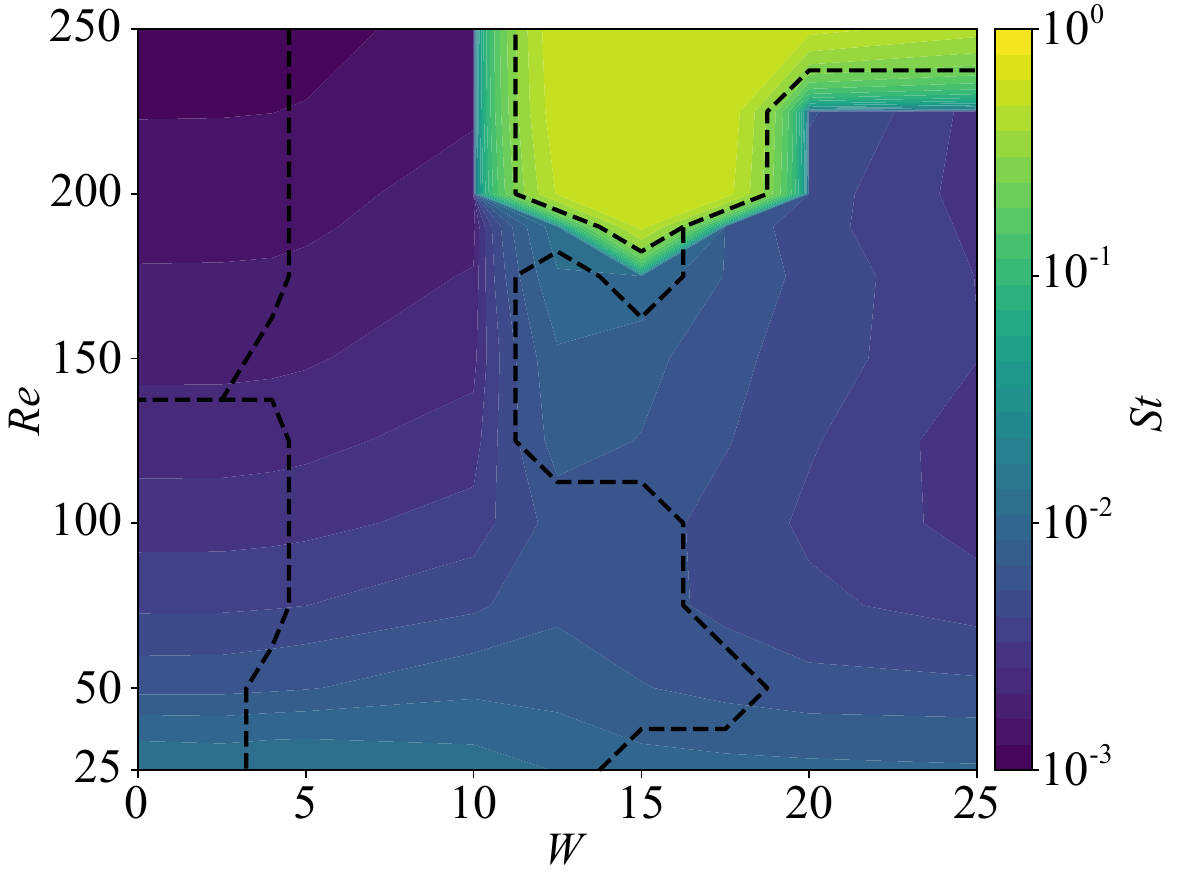}
    \caption{}
\end{subfigure}
  \caption{Thermal performance factor $\eta$ and Stanton number $St$ contours corresponding to the $Re$--$W$ parameter space in Figure~\ref{fig:nu_heatmap}(a). The black dashed lines correspond to the boundaries between different flow regimes in Figure~\ref{fig:heatmapPKE}.}
\label{fig:engineering}
\end{figure}

While the HTE quantified by the Nusselt number $\langle Nu \rangle$ is substantial, it inherently considers heat transfer in the vertical direction. In this section, we also consider the thermal performance associated with the through-flow to give a comprehensive evaluation of the RBP flow model from an engineering perspective.
Specifically, we introduce the thermal performance factor $\eta$ \citep{maradiya_heat_2018} to characterise the thermal-hydraulic efficiency and the Stanton number $St$ \citep{takahashi_experimental_1992} to characterise the heat transfer intensity.

The thermal performance factor is defined as:
\begin{equation}
\eta =
\frac{
\langle Nu\rangle_{W>0}/\langle Nu\rangle_{W=0}
}{
\left(f_{W>0}/f_{W=0}\right)^{1/3}
},
\end{equation}
where $f \propto {\sigma_w}/U_{bulk}^2$ represents the friction factor, with $\sigma_w$ denoting the wall shear stress. If $\eta > 1$, the benefits of HTE outweigh the additional pumping power requirements, making the design energy-efficient. Conversely, $\eta < 1$ implies that the frictional losses dominate, and $\eta = 1$ represents a break-even scenario. Since the present simulations are driven by a constant pressure gradient, the friction drag on the channel wall remains constant ($\sigma_{w,W>0} = \sigma_{w,W=0}$). Consequently, the thermal performance factor simplifies to:
\begin{equation}
\eta  = \frac{{\left\langle {Nu} \right\rangle }_{W>0}}{{\left\langle {Nu} \right\rangle }_{W=0}}{\left( {\frac{{{U_{bulk,W>0}}}}{{{U_{bulk,W=0}}}}} \right)^{\frac{2}{3}}}.
\end{equation}

The Stanton number is defined as:  
\begin{equation}
    St = \frac{{\left\langle {Nu} \right\rangle }}{Pr Re {{U_{bulk}}}},
\end{equation}
where a large $St$ indicates rapid thermal equilibration with the boundary, and hence short characteristic streamwise and temporal scales for heat uptake.

The results over the same $Re$--$W$ parameter space as that shown in Figure~\ref{fig:nu_heatmap}(a) are presented in Figure~\ref{fig:engineering}. The upper blue region in Figure~\ref{fig:engineering}(a), corresponding to $\eta \approx 0.5$, and the upper green region in Figure~\ref{fig:engineering}(b), corresponding to $St \approx 0.6$, coincide with the significant $Nu$ and the wall-attached hook structure in Figure~\ref{fig:nu_heatmap}(a). The relatively low value of $\eta$ indicates that the additional pumping power required increases more rapidly than the overall increase of HTE, rendering this regime energetically inefficient. By contrast, the exceptionally large value of $St \approx 0.6$, which is over $500$ times the corresponding Newtonian value, implies that the flow can reach thermal equilibrium with the channel walls over very short temporal and spatial scales. This regime therefore appears promising for process-stream thermal conditioning that requires rapid thermal equilibration, such as polymer extrusion \citep{ali_effects_2016, khan_mhd_2017}.

In Figure~\ref{fig:engineering}(a), a red region with $\eta > 1$ is also identified, making this regime attractive for heat-transport applications, such as cooling systems. In this regime, the nonlinear state takes the form of a wall-detached hook and may arise as either a travelling wave or periodic orbits. The maximum thermal performance factor $\eta = 1.385$ is attained at $Re = 175$ and $W = 12.5$, for which $\langle Nu \rangle = 3.452$, representing an increase of approximately $150\%$ relative to the Newtonian value $\langle Nu \rangle = 1.351$. The corresponding Stanton number is $St = 1.055 \times 10^{-2}$, which is approximately 6 times the Newtonian value $St = 1.648 \times 10^{-3}$.

To assess the practical applicability of these results, we place the two operating conditions discussed above in a dimensional context. Table~\ref{tab:dimensional_config} presents representative configurations for water-based systems with a fixed temperature difference of $\Delta\theta^*=50^\circ\mathrm{C}$. Case~1 corresponds to the maximum thermal performance factor $\eta$ and hence the greatest thermal-hydraulic efficiency identified, whereas case~2 corresponds to the largest HTE obtained, reaching approximately $1100\%$. The resulting channel half-heights, $H=6.3$ and $12.6~\mathrm{mm}$, and reference velocities, $U^*=0.0280$ and $0.0397~\mathrm{ms^{-1}}$, respectively, are readily attainable experimentally, providing practical benchmark configurations for validating the proposed polymer-enhanced heat-transfer mechanism and guiding its translation to microchannel-related applications.

\begin{table}
  \centering
  \begin{tabular}{ccccccccc}
    \toprule
    Case & $Re$ & $W$  &$\beta$& $Ri$   & $Pr$  &         HTE      & $H$$(\mathrm{m})$ & $U^*$$(\mathrm{ms^{-1}})$  \\
    \midrule
     1   & 175  & 12.5 & 0.98  & $-0.8$ &  7 & $\approx 150\%$  &  0.0063 & 0.0280   \\
     2   & 500  & 12.5 & 0.98  & $-0.8$ &  7 &$\approx 1100\%$ &  0.0126  & 0.0397   \\
    \bottomrule
  \end{tabular}
\caption{Representative dimensional configurations for a fixed temperature difference $\Delta\theta^*=50^\circ\mathrm{C}$. The dimensionalisation uses a thermal diffusivity $\kappa=1.43\times10^{-7}$~$\mathrm{m^2s^{-1}}$, coefficient of thermal expansion $\alpha=2\times10^{-4}$~$\mathrm{K^{-1}}$ , kinematic viscosity $\nu=10^{-6}$~$\mathrm{m^2s^{-1}}$, gravitational acceleration $g=9.8$~$\mathrm{ms^{-2}}$.}
  \label{tab:dimensional_config}
\end{table}

Finally, it is worth re-iterating that only 2D results have been presented here in which the relevant convective instability consists of transverse rolls. In 3D, longitudinal roll solutions can be more unstable depending on the aspect ratio of the channel \citep[e.g.][]{Nicolas2000} and understanding what the effect of viscoelasticity on these remains to be explored. Certainly for a tall, thin channel, the 2D dynamics discussed here should still be relevant.

%
%
\section{Conclusions}
\label{sec:conclusions}

In this paper, we investigated the effects of adding polymer additives on the heat transfer possible in 2D stratified flows using 
linear stability analysis and nonlinear DNS. In the linear stability analysis, two distinct modes were observed: the convective mode and the elastic centre mode. Elasticity alters the convective mode primarily through the velocity eigenfunctions $\tilde{u}$ and $\tilde{w}$, in which sharp gradients develop in the vicinity of $z=\pm 0.5$. In contrast, the eigenfunctions associated with the centre mode remain largely unaffected by stratification. However, stratification modifies the growth rate of this mode, with the growth rate decreasing as $Ri$ increases. The convective mode is likewise suppressed with increasing $Ri$, but it exhibits a stronger sensitivity to stratification than the centre mode.

DNS reveals that the elastic mode (which gives rise to a nonlinear arrowhead state) produces negligible enhanced heat transfer (only 0.030\%) (Figure~\ref{fig:centre_cases}) due to its weak vertical velocities. In contrast, a viscoelastically-modified convective mode can produce significantly more heat transport. The nonlinear resulting state triggered by the convective mode manifests as either periodic orbits or travelling waves. Both states are characterised by \emph{hook-like} polymer-stress structures that act as effective `polymer walls' to steer the flow along the polymer-stretching directions and increase vertical heat transport. An exploration of $Re$--$W$ space reveals that a pronounced HTE occurs at $W \in \left[ 10,15 \right]$ for all sufficiently large $Re$ (Figure~\ref{fig:nu_heatmap}(a)). This HTE regime is closely associated with the form of the hook structures (Figure~\ref{fig:relative_flow_field}) and their propagation speed (Figure~\ref{fig:nu_vs_cp_uc}). These structures remain detached from the channel walls at lower $Re$. In this regime, they act as walls in the channel interior, deflecting fluid vertically around them and reducing the flow rate, which increases $\langle Nu \rangle$. The hook structures migrate outward toward the walls as $Re$ or $W$ increases, increasing their magnitude and causing  $\langle Nu \rangle$ to increase further.

The abrupt amplification in HTE occurs when $Re$ exceeds a critical value and the hook structures attach to the wall. The structures appear as travelling-wave solutions that move much more slowly than the detached hooks. They likewise act as effective walls, reorganising the flow into strong counter-rotating rolls that markedly enhance heat transfer but also induce a sharp reduction in the flow rate.  The largest enhancement is attained in this regime at $Re = 500$ and $W = 12.5$, where $\langle Nu \rangle$ is 1100\% greater than its Newtonian value (Figure~\ref{fig:nu_heatmap}(b)); this enhancement may increase further at higher $Re$. The smooth evolution from the wall-detached to the wall-attached hook structure suggests that these are two manifestations of the same underlying flow structure (Figure~\ref{fig:tr_evolution}).

The perturbation kinetic energy budget analysis reveals the elasto-buoyant nature of the travelling-wave and periodic-orbit states. These nonlinear states are sustained by the interplay between elastic and buoyancy forces, and the most pronounced HTE occurs when the polymer-work–buoyancy-flux ratio satisfies $\mathcal{T}/\mathcal{B}>0$, indicating that both contributions act to destabilise the flow relative to the laminar state.
The ratio $\mathcal{T}/\mathcal{B}$ increases monotonically with increasing $W$, suggesting a progressively more dominant elastic effect. Beyond a critical value of $W$, the nonlinear state undergoes a transition from a travelling wave to a periodic orbit state. This dependence on $W$ suggests that the periodic orbits are sustained primarily by elastic forces. As $W$ increases further, the flow enters a relaminarisation-cycle regime. In this regime, the cycle period scales linearly with the polymer relaxation time, indicating that polymer stresses have fully overtaken buoyancy as the dominant dynamical influence.

The practical implications of the through-flow are evaluated by thermal performance from an engineering perspective.
The results (Figure~\ref{fig:engineering}) reveal two distinct operating regimes. In the wall-attached hook regime, both the HTE and the Stanton number $St$ are extremely large, indicating strong vertical heat transport and rapid thermal equilibration with the walls, albeit at the expense of a substantial hydraulic penalty ($\eta < 1$). This regime is therefore potentially attractive for thermal-conditioning processes such as polymer extrusion. By contrast, when the hook structure remains detached from the walls, the system can achieve substantial HTE while maintaining $\eta > 1$, making this regime more favourable for energy-efficient heat-transport applications. The most thermally efficient operating condition identified here occurs at $Re = 175$ and $W = 12.5$, where $\eta = 1.385$, the heat transfer is enhanced by $150\%$ relative to the Newtonian baseline, and $St$ is about $6$ times its Newtonian value.

In terms of future work, the fact that such polymer-enhanced heat flux can be found in our 2D system strongly motivates a more extensive (and computationally expensive) 3D study. After all, Squire's theorem does not hold over almost the entire range of unstable stratification \citep{gage_stability_1968, kelly_onset_1994}  and so (convective) longitudinal rolls can be more unstable than the (transverse roll) convective mode considered here in a wide enough domain. If this longitudinal roll instability does not lead to similar heat transfer enhancement, it may still be possible to engineer the 2D (transverse roll) scenario studied here by designing thin enough cooling channels \citep[e.g.][]{Nicolas2000}. We hope to report on this important extension in the near future.

\begin{bmhead}[Acknowledgements.]
Some of the simulations were carried out on the CSD3 platform at the University of Cambridge. The authors gratefully acknowledge initial funding from the MPhil in Energy Technologies programme in the Department of Engineering for this work.
\end{bmhead}

\begin{bmhead}[Declaration of interests.]
The authors report no conflict of interest.
\end{bmhead}

\appendix

\section{The fluctuation kinetic energy (FKE) budget}
\label{app:fke_budget}

In this section, we perform the classical fluctuation kinetic energy (FKE) budget following \cite{pope_turbulent_2001} and compare the result with the PKE budget analysis in \S\ref{sec:energy_budget}.
For the FKE budget, we first reconstruct the total field $f_i = F_i + f_i'$ according to \eqref{eq:base_state_decomposition} and subsequently decompose it into average and fluctuating components:
\[
{f_i} = \left[\kern-0.15em\left[ {{f_i}} 
 \right]\kern-0.15em\right]_z + {f_{i}''}.
 \]

Both FKE and PKE budgets can be cast into a unified equation of the following form:
\begin{equation}
    \frac{\partial k^\alpha}{\partial t} + \partial_j \mathcal{F}_j^\alpha
    = \mathcal{P}^\alpha + \mathcal{B}^\alpha + \mathcal{T}^\alpha + \mathcal{D}^\alpha,
    \label{eq:tkepke_equation}
\end{equation}
where the superscript $\alpha  \in \left\{ {f,p} \right\}$ denotes either the FKE ($f$) or PKE ($p$) budget.

%
%
\begin{figure}
  \centerline{\includegraphics[width=0.6\linewidth]{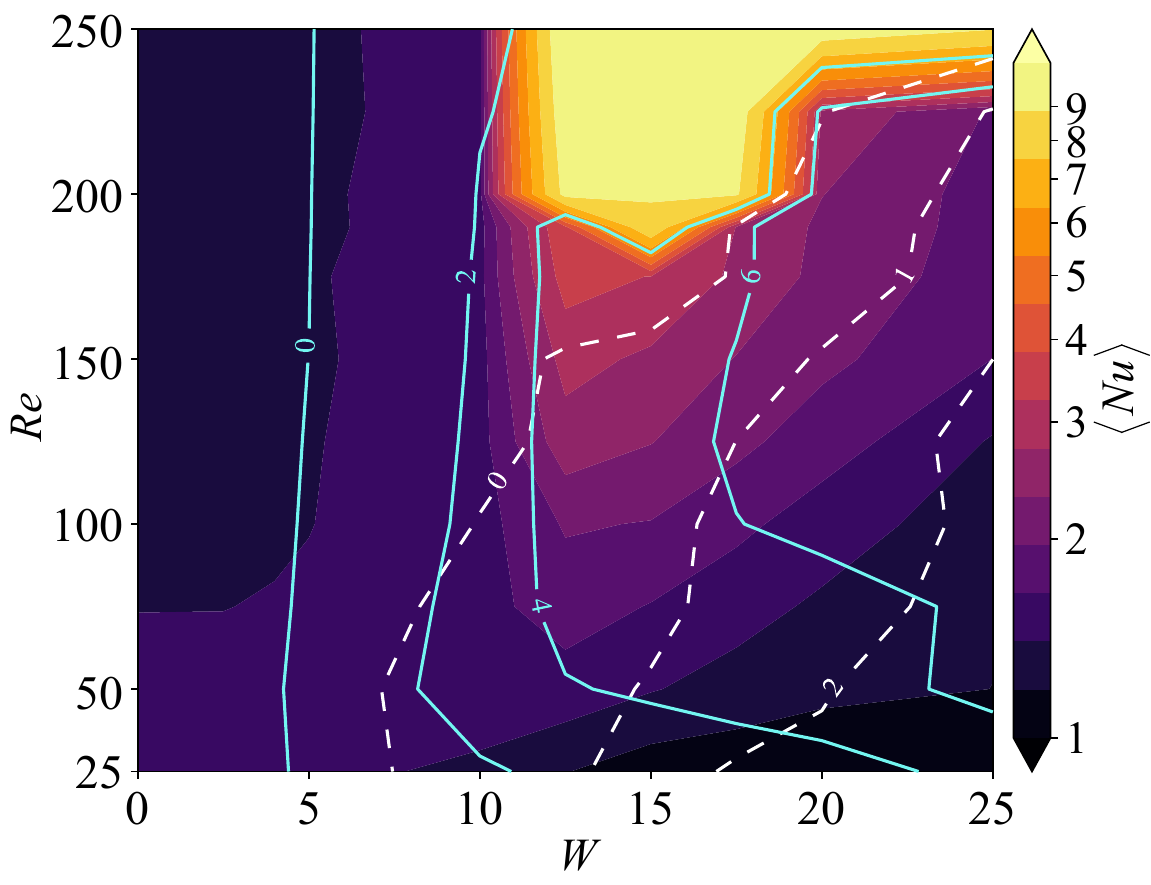}}
  \caption{Contours of $\langle Nu\rangle$ in the $Re$--$W$ parameter space with FKE and PKE budget terms for fixed $\beta = 0.98$ and $Ri = -0.8$. The blue contour lines represent the ratio of polymer conversion to buoyancy flux, $\mathcal{T}^f/\mathcal{B}^f$ (blue dashed line) and $\mathcal{T}^p/\mathcal{B}^p$ (blue solid line), as defined in equation \eqref{eq:tkepke_equation}.}
  \label{fig:heatmap_and_TKEFKE}
\end{figure}

The distinction between the FKE and PKE budgets highlights the different physical mechanisms captured by the two decomposition frameworks. The FKE budget is based on time averaging, where the reference state is the time-averaged mean flow; thus, it captures the energy of fluctuations relative to the statistical mean. Conversely, the PKE budget is defined relative to the laminar base state, measuring the energy associated with the deviation from the laminar solution.


%
In Figure~\ref{fig:heatmap_and_TKEFKE}, the production terms ($\mathcal{P}^f$, $\mathcal{P}^p$) remain consistently negative and the buoyancy flux terms ($\mathcal{B}^f$, $\mathcal{B}^p$) remain consistently positive across the entire $Re$--$W$ parameter space.
Although the FKE budget shows no direct correspondence with the HTE phenomenon, it nevertheless reveals a change in the role of polymer elasticity as $W$ varies.
Both $\mathcal{T}^f/\mathcal{B}^f$ and $\mathcal{T}^p/\mathcal{B}^p$ ratios increase monotonically with increasing $W$. In the small $W$ regime (between these two zero boundaries), $\mathcal{T}^f/\mathcal{B}^f < 0$ while $\mathcal{T}^p/\mathcal{B}^p > 0$. This indicates that polymers suppress fluctuating motions but promote deviation from the base state. 
However, at larger $W$ (in the regime to the right of $\mathcal{T}^f/\mathcal{B}^f = 0$), the polymer conversion terms become positive in both the FKE and PKE analyses. This indicates that polymers shift to promoting fluctuations, continuously driving the flow away from the base state.


\phantomsection
\addcontentsline{toc}{section}{References}
\bibliographystyle{jfm}
\bibliography{references}

@article{khalid_centre-mode_2021,
  author  = {Khalid, M. and Chaudhary, I. and Garg, P. and Shankar, V. and Subramanian, G.},
  title   = {The centre-mode instability of viscoelastic plane {Poiseuille} flow},
  journal = {J. Fluid Mech.},
  volume  = {915},
  pages   = {A43},
  year    = {2021},
  doi     = {10.1017/jfm.2021.60}
}

@article{gage_stability_1968,
  author  = {Gage, K. S. and Reid, W. H.},
  title   = {The stability of thermally stratified plane {Poiseuille} flow},
  journal = {J. Fluid Mech.},
  volume  = {33},
  pages   = {21--32},
  year    = {1968},
  doi     = {10.1017/S0022112068002326}
}

@article{sureshkumar_linear_1995,
  author  = {Sureshkumar, R. and Beris, A. N.},
  title   = {Linear stability analysis of viscoelastic {Poiseuille} flow using an {Arnoldi}-based orthogonalization algorithm},
  journal = {J. Non-Newtonian Fluid Mech.},
  volume  = {56},
  pages   = {151--182},
  year    = {1995},
  doi     = {10.1016/0377-0257(94)01279-Q}
}

@incollection{kelly_onset_1994,
  author    = {Kelly, R. E.},
  title     = {The onset and development of thermal convection in fully developed shear flows},
  booktitle = {Advances in Applied Mechanics},
  editor    = {Hutchinson, J. W. and Wu, T. Y.},
  volume    = {31},
  pages     = {35--112},
  publisher = {Elsevier},
  year      = {1994},
  doi       = {10.1016/S0065-2156(08)70255-2}
}

@article{yao_effects_2024,
  author  = {Yao, Z.-Z. and Lu, C.-L. and Zhou, C.-T. and Luo, K. and Yi, H.-L. and Tan, H.-P.},
  title   = {Effects of shear intensity on the linear instability of viscoelastic {Rayleigh--B{\'e}nard--Poiseuille} flow},
  journal = {Intl J. Heat Fluid Flow},
  volume  = {107},
  pages   = {109336},
  year    = {2024},
  doi     = {10.1016/j.ijheatfluidflow.2024.109336}
}

@article{burns_dedalus_2020,
  author  = {Burns, K. J. and Vasil, G. M. and Oishi, J. S. and Lecoanet, D. and Brown, B. P.},
  title   = {{Dedalus}: a flexible framework for numerical simulations with spectral methods},
  journal = {Phys. Rev. Research},
  volume  = {2},
  pages   = {023068},
  year    = {2020},
  doi     = {10.1103/PhysRevResearch.2.023068}
}

@article{dubief_first_2022,
  author  = {Dubief, Y. and Page, J. and Kerswell, R. R. and Terrapon, V. E. and Steinberg, V.},
  title   = {First coherent structure in elasto-inertial turbulence},
  journal = {Phys. Rev. Fluids},
  volume  = {7},
  pages   = {073301},
  year    = {2022},
  doi     = {10.1103/PhysRevFluids.7.073301}
}

@article{dong_asymptotic-analysis-inspired_2025,
  author        = {Dong, M. and Wan, D.},
  title         = {Asymptotic-analysis-inspired boundary conditions aiming at eliminating polymer diffusive instability},
  year          = {2025},
  journal       = {J. Fluid Mech.},
  pages         = {A8},
  volume        = {1020}
}

@article{beneitez_polymer_2023,
  author  = {Beneitez, M. and Page, J. and Kerswell, R. R.},
  title   = {Polymer diffusive instability leading to elastic turbulence in plane {Couette} flow},
  journal = {Phys. Rev. Fluids},
  volume  = {8},
  pages   = {L101901},
  year    = {2023},
  doi     = {10.1103/PhysRevFluids.8.L101901}
}

@article{zhang_review_2021,
  author  = {Zhang, Z. and Wang, X. and Yan, Y.},
  title   = {A review of the state-of-the-art in electronic cooling},
  journal = {e-Prime},
  volume  = {1},
  pages   = {100009},
  year    = {2021},
  doi     = {10.1016/j.prime.2021.100009}
}

@article{konovalov_recent_2023,
  author  = {Konovalov, D. and Tolstorebrov, I. and Eikevik, T. M. and Kobalava, H. and Radchenko, M. and Hafner, A. and Radchenko, A.},
  title   = {Recent developments in cooling systems and cooling management for electric motors},
  journal = {Energies},
  volume  = {16},
  pages   = {7006},
  year    = {2023},
  doi     = {10.3390/en16197006}
}

@article{hirata_convective_2015,
  author  = {Hirata, S. C. and Alves, L. S. de B. and Delenda, N. and Ouarzazi, M. N.},
  title   = {Convective and absolute instabilities in {Rayleigh--B{\'e}nard--Poiseuille} mixed convection for viscoelastic fluids},
  journal = {J. Fluid Mech.},
  volume  = {765},
  pages   = {167--210},
  year    = {2015},
  doi     = {10.1017/jfm.2014.721}
}

@article{roy_viscoelastic_2024,
  author  = {Roy, T. and Taylor, D. and Poulikakos, D.},
  title   = {Viscoelastic flow instabilities for enhanced heat transfer in battery pack cooling},
  journal = {Intl J. Heat Mass Transfer},
  volume  = {232},
  pages   = {125888},
  year    = {2024},
  doi     = {10.1016/j.ijheatmasstransfer.2024.125888}
}

@book{schmid_stability_2001,
  author    = {Schmid, P. J. and Henningson, D. S.},
  title     = {Stability and Transition in Shear Flows},
  publisher = {Springer},
  address   = {New York},
  year      = {2001},
  doi       = {10.1007/978-1-4613-0185-1}
}

@article{nicolas_revue_2002,
  author  = {Nicolas, X.},
  title   = {Revue bibliographique sur les {\'e}coulements de {Poiseuille--Rayleigh--B{\'e}nard}: {\'e}coulements de convection mixte en conduites rectangulaires horizontales chauff{\'e}es par le bas},
  journal = {Intl J. Thermal Sci.},
  volume  = {41},
  pages   = {961--1016},
  year    = {2002},
  doi     = {10.1016/S1290-0729(02)01374-1}
}

@article{groisman_elastic_2000,
  author  = {Groisman, A. and Steinberg, V.},
  title   = {Elastic turbulence in a polymer solution flow},
  journal = {Nature},
  volume  = {405},
  pages   = {53--55},
  year    = {2000},
  doi     = {10.1038/35011019}
}

@book{pope_turbulent_2001,
  author    = {Pope, S. B.},
  title     = {Turbulent Flows},
  publisher = {Cambridge University Press},
  address   = {Cambridge},
  year      = {2000}
}

@article{Page_2020,
  author  = {Page, J. and Dubief, Y. and Kerswell, R. R.},
  title   = {Exact traveling wave solutions in viscoelastic channel flow},
  journal = {Phys. Rev. Lett.},
  volume  = {125},
  pages   = {154501},
  year    = {2020},
  doi     = {10.1103/PhysRevLett.125.154501}
}

@article{wu_cooling_2025,
  author  = {Wu, Z. and Wang, Z.},
  title   = {Cooling next-generation electronics: from emerging semiconductors to data centers},
  journal = {The Innovation Energy},
  volume  = {2},
  pages   = {100125},
  year    = {2025},
  doi     = {10.59717/j.xinn-energy.2025.100125}
}

@article{abed_experimental_2016,
  author  = {Abed, W. M. and Whalley, R. D. and Dennis, D. J. C. and Poole, R. J.},
  title   = {Experimental investigation of the impact of elastic turbulence on heat transfer in a serpentine channel},
  journal = {J. Non-Newtonian Fluid Mech.},
  volume  = {231},
  pages   = {68--78},
  year    = {2016},
  doi     = {10.1016/j.jnnfm.2016.03.003}
}

@article{whalley_enhancing_2015,
  author  = {Whalley, R. D. and Abed, W. M. and Dennis, D. J. C. and Poole, R. J.},
  title   = {Enhancing heat transfer at the micro-scale using elastic turbulence},
  journal = {Theor. Appl. Mech. Lett.},
  volume  = {5},
  pages   = {103--106},
  year    = {2015},
  doi     = {10.1016/j.taml.2015.03.006}
}

@article{li_numerical_2017,
  author  = {Li, D.-Y. and Zhang, H. and Cheng, J.-P. and Li, X.-B. and Li, F.-C. and Qian, S. and Joo, S. W.},
  title   = {Numerical simulation of heat transfer enhancement by elastic turbulence in a curvy channel},
  journal = {Microfluid. Nanofluid.},
  volume  = {21},
  pages   = {25},
  year    = {2017},
  doi     = {10.1007/s10404-017-1859-x}
}

@article{yang_experimental_2020,
  author  = {Yang, H. and Yao, G. and Wen, D.},
  title   = {Experimental investigation on convective heat transfer of shear-thinning fluids by elastic turbulence in a serpentine channel},
  journal = {Exp. Thermal Fluid Sci.},
  volume  = {112},
  pages   = {109997},
  year    = {2020},
  doi     = {10.1016/j.expthermflusci.2019.109997}
}

@article{traore_efficient_2015,
  author  = {Traore, B. and Castelain, C. and Burghelea, T.},
  title   = {Efficient heat transfer in a regime of elastic turbulence},
  journal = {J. Non-Newtonian Fluid Mech.},
  volume  = {223},
  pages   = {62--76},
  year    = {2015},
  doi     = {10.1016/j.jnnfm.2015.05.005}
}

@article{garg_enhanced_2024,
  author  = {Garg, H. and Wang, L.},
  title   = {Enhanced heat transfer in a two-dimensional serpentine micro-channel using elastic polymers},
  journal = {Intl J. Thermofluids},
  volume  = {23},
  pages   = {100724},
  year    = {2024},
  doi     = {10.1016/j.ijft.2024.100724}
}

@article{cai_polymer_2019,
  author  = {Cai, W. and Wei, T. and Tang, X. and Liu, Y. and Li, B. and Li, F.},
  title   = {The polymer effect on turbulent {Rayleigh--B{\'e}nard} convection based on {PIV} experiments},
  journal = {Exp. Thermal Fluid Sci.},
  volume  = {103},
  pages   = {214--221},
  year    = {2019},
  doi     = {10.1016/j.expthermflusci.2019.01.011}
}

@article{wang_pattern_2023,
  author  = {Wang, Y. and Cheng, J.-P. and Zhang, H.-N. and Zheng, X. and Cai, W.-H. and Siginer, D. A.},
  title   = {Pattern selection and heat transfer in the {Rayleigh--B{\'e}nard} convection near the vicinity of the convection onset with viscoelastic fluids},
  journal = {Phys. Fluids},
  volume  = {35},
  pages   = {013104},
  year    = {2023},
  doi     = {10.1063/5.0132949}
}

@article{xu_direct_2025,
  author  = {Xu, C. and Zhang, C. and Brandt, L. and Song, J. and Shishkina, O.},
  title   = {Direct numerical simulations of turbulent {Rayleigh--B{\'e}nard} convection with polymer additives},
  journal = {J. Fluid Mech.},
  volume  = {1014},
  pages   = {A22},
  year    = {2025},
  doi     = {10.1017/jfm.2025.10286}
}

@article{sharma_energy_2015,
  author  = {Sharma, C. S. and Tiwari, M. K. and Zimmermann, S. and Brunschwiler, T. and Schlottig, G. and Michel, B. and Poulikakos, D.},
  title   = {Energy efficient hotspot-targeted embedded liquid cooling of electronics},
  journal = {Appl. Energy},
  volume  = {138},
  pages   = {414--422},
  year    = {2015},
  doi     = {10.1016/j.apenergy.2014.10.068}
}

@article{khalili_hybrid_2023,
  author  = {Khalili, Z. and Sheikholeslami, M. and Momayez, L.},
  title   = {Hybrid nanofluid flow within cooling tube of photovoltaic-thermoelectric solar unit},
  journal = {Sci. Rep.},
  volume  = {13},
  pages   = {8202},
  year    = {2023},
  doi     = {10.1038/s41598-023-35428-6}
}

@article{taher_experimental_2018,
  author  = {Taher, R. and Abid, C.},
  title   = {Experimental determination of heat transfer in a {Poiseuille--Rayleigh--B{\'e}nard} flow},
  journal = {Heat Mass Transfer},
  volume  = {54},
  pages   = {1453--1466},
  year    = {2018},
  doi     = {10.1007/s00231-017-2220-3}
}

@article{taher_poiseuille-rayleigh-benard_2021,
  author  = {Taher, R. and Ahmed, M. M. and Haddad, Z. and Abid, C.},
  title   = {{Poiseuille--Rayleigh--B{\'e}nard} mixed convection flow in a channel: heat transfer and fluid flow patterns},
  journal = {Intl J. Heat Mass Transfer},
  volume  = {180},
  pages   = {121745},
  year    = {2021},
  doi     = {10.1016/j.ijheatmasstransfer.2021.121745}
}

@article{zhang_three-dimensional_2023,
  author  = {Zhang, L. and Huang, Y. and Peng, L. and Li, Y.-R.},
  title   = {Three-dimensional numerical simulations on {Poiseuille--Rayleigh--B{\'e}nard} convection of air in a horizontal rectangular channel},
  journal = {Case Stud. Thermal Engng},
  volume  = {49},
  pages   = {103309},
  year    = {2023},
  doi     = {10.1016/j.csite.2023.103309}
}

@article{yan_enhanced_2023,
  author  = {Yan, R. and Wu, B. and Gao, C. and Li, Y.},
  title   = {Enhanced heat transfer in {Poiseuille--Rayleigh--B{\'e}nard} flows based on dielectric-barrier-discharge plasma actuation},
  journal = {Phys. Plasmas},
  volume  = {30},
  pages   = {033501},
  year    = {2023},
  doi     = {10.1063/5.0131414}
}

@article{yan_study_2026,
  author  = {Yan, R. and Gao, C. and Zheng, Y. and Wu, B. and Zheng, H. and Ding, T.},
  title   = {A study on the mechanism of heat transfer enhancement in {Poiseuille--Rayleigh--B{\'e}nard} flow using {SDBD} plasma actuation},
  journal = {Intl Commun. Heat Mass Transfer},
  volume  = {171},
  pages   = {110084},
  year    = {2026},
  doi     = {10.1016/j.icheatmasstransfer.2025.110084}
}

@article{yao_non-modal_2025,
  author  = {Yao, Z. and Lu, C. and Zhang, M. and Luo, K. and Yi, H.},
  title   = {Non-modal analysis of {Rayleigh--B{\'e}nard} convection with and without bounded shear flows in viscoelastic fluids},
  journal = {J. Fluid Mech.},
  volume  = {1011},
  pages   = {A37},
  year    = {2025},
  doi     = {10.1017/jfm.2025.291}
}

@article{kamboj_exacerbation_2025,
  author  = {Kamboj, A. and Patne, R. and Narayana, P. A. L. and Sahu, K. C.},
  title   = {Exacerbation of viscoelastic instability due to viscous heating},
  journal = {Chem. Engng Sci.},
  volume  = {316},
  pages   = {121989},
  year    = {2025},
  doi     = {10.1016/j.ces.2025.121989}
}

@article{zhang_linear_2013,
  author  = {Zhang, M. and Lashgari, I. and Zaki, T. A. and Brandt, L.},
  title   = {Linear stability analysis of channel flow of viscoelastic {Oldroyd-B} and {FENE-P} fluids},
  journal = {J. Fluid Mech.},
  volume  = {737},
  pages   = {249--279},
  year    = {2013},
  doi     = {10.1017/jfm.2013.572}
}

@article{dubief2023elasto,
  author  = {Dubief, Y. and Terrapon, V. E. and Hof, B.},
  title   = {Elasto-inertial turbulence},
  journal = {Annu. Rev. Fluid Mech.},
  volume  = {55},
  pages   = {675--705},
  year    = {2023},
  doi     = {10.1146/annurev-fluid-032822-025933}
}

@article{steinberg2021elastic,
  author  = {Steinberg, V.},
  title   = {Elastic turbulence: an experimental view on inertialess random flow},
  journal = {Annu. Rev. Fluid Mech.},
  volume  = {53},
  pages   = {27--58},
  year    = {2021},
  doi     = {10.1146/annurev-fluid-010719-060129}
}

@article{Miles61,
  author  = {Miles, J. W.},
  title   = {On the stability of heterogeneous shear flows},
  journal = {J. Fluid Mech.},
  volume  = {10},
  pages   = {496--508},
  year    = {1961},
  doi     = {10.1017/S0022112061000305}
}

@article{Howard61,
  author  = {Howard, L. N.},
  title   = {Note on a paper of {John W. Miles}},
  journal = {J. Fluid Mech.},
  volume  = {10},
  pages   = {509--512},
  year    = {1961},
  doi     = {10.1017/S0022112061000317}
}

@article{LewyK25,
  author        = {Lewy, T. A. and Kerswell, R. R.},
  title         = {Localised arrowheads: the building blocks of elastic turbulence in rectilinear, sheared polymer flows},
  journal       = {Phys. Rev. Fluids},
  volume        = {11},
  pages         = {L051301},
  year          = {2026}
}

@article{maradiya_heat_2018,
  author  = {Maradiya, C. and Vadher, J. and Agarwal, R.},
  title   = {The heat transfer enhancement techniques and their thermal performance factor},
  journal = {Beni-Suef Univ. J. Basic Appl. Sci.},
  volume  = {7},
  pages   = {1--21},
  year    = {2018},
  doi     = {10.1016/j.bjbas.2017.10.001}
}

@incollection{manglik_heat_2003,
  author    = {Manglik, R. M.},
  title     = {Heat transfer enhancement},
  booktitle = {Heat Transfer Handbook},
  publisher = {Wiley},
  address   = {New York},
  pages     = {1029--1130},
  year      = {2003}
}

@article{ali_effects_2016,
  author  = {Ali, N. and Javed, M. A.},
  title   = {Effects of temperature-dependent properties on wire-coating from a bath of {FENE-P} fluid},
  journal = {Intl J. Heat Mass Transfer},
  volume  = {103},
  pages   = {401--410},
  year    = {2016},
  doi     = {10.1016/j.ijheatmasstransfer.2016.07.072}
}

@article{khan_mhd_2017,
  author  = {Khan, Z. and Shah, R. A. and Islam, S. and Jan, H. and Jan, B. and Rasheed, H.-U. and Khan, A.},
  title   = {{MHD} flow and heat transfer analysis in the wire coating process using elastic-viscous fluid},
  journal = {Coatings},
  volume  = {7},
  pages   = {15},
  year    = {2017},
  doi     = {10.3390/coatings7010015}
}

@article{lin_weakening_2025,
  author  = {Lin, Y. and Kerswell, R. R.},
  title   = {Weakening the effect of boundaries: diffusion-free boundary conditions as a do least harm alternative to {Neumann}},
  journal = {Geophys. Astrophys. Fluid Dyn.},
  volume  = {119},
  pages   = {525--549},
  year    = {2025},
  doi     = {10.1080/03091929.2024.2403759}
}

@article{takahashi_experimental_1992,
  author  = {Takahashi, M. and Fujinuma, H. and Tsukuii, J. and Inoue, A.},
  title   = {Experimental study on condensation heat transfer on surface of liquid jet},
  journal = {J. Nucl. Sci. Technol.},
  volume  = {29},
  pages   = {721--734},
  year    = {1992},
  doi     = {10.1080/18811248.1992.9731588}
}

@article{dubief_heat_2020,
  author  = {Dubief, Y. and Terrapon, V. E.},
  title   = {Heat transfer enhancement and reduction in low-{Rayleigh} number natural convection flow with polymer additives},
  journal = {Phys. Fluids},
  volume  = {32},
  pages   = {033103},
  year    = {2020},
  doi     = {10.1063/1.5143275}
}

@article{Datta22,
  title={Perspectives on viscoelastic flow instabilities and elastic turbulence},
  author={Datta, Sujit S and Ardekani, Arezoo M and Arratia, Paulo E and Beris, Antony N and Bischofberger, Irmgard and McKinley, Gareth H and Eggers, Jens G and L{\'o}pez-Aguilar, J Esteban and Fielding, Suzanne M and Frishman, Anna and others},
  journal={Phys. Rev. Fluids},
  volume={7},
  number={8},
  pages={080701},
  year={2022},
  publisher={APS}
}

@article{Castillo22,
  title={Understanding viscoelastic flow instabilities: {O}ldroyd-{B} and beyond},
  author={S{\'a}nchez, Hugo A Castillo and Jovanovi{\'c}, Mihailo R and Kumar, Satish and Morozov, Alexander and Shankar, V and Subramanian, Ganesh and Wilson, Helen J},
  journal={J. Non-Newton. Fluid Mech.},
  pages={104742},
  year={2022},
  publisher={Elsevier}
}

@article{Serafini22,
  title={Drag reduction in turbulent wall-bounded flows of realistic polymer solutions},
  author={Serafini, F. and Battista, F. and Gualtieri, P. and Casciola, C.M.},
  journal={Phys. Rev. Lett.},
  pages={104502},
  volume={129},
  year={2022},
}

@article{Shaqfeh21,
  title={The Oldroyd-B fluid in elastic instabilities, turbulence and particle suspensions},
  author={Shaqfeh, E. S. G. and Khomami, B.},
  journal={J. Non-Newton. Fluid Mech.},
  pages={104672},
  volume={298},
  year={2021},
}

@article{Nicolas2000,
  title={Linear stability of mixed convection flows in horizontal rectangular channels of finite transversal extension heated from below},
  author={Nicolas, X. and Luijkx, J.-M. and Platten, J.-K.},
  journal={Int. J. Heat Mass Transfer},
  pages={589-610},
  volume={43},
  year={2000},
}

\end{document}